\documentclass[12pt,preprint2]{aastex}
\input epsf.sty

\shorttitle{Radiation pressure instability}
\shortauthors{Janiuk et al.}

\begin{document}

\title{Radiation pressure instability driven variability in
the accreting black holes}

\author{Agnieszka Janiuk and Bo\.zena Czerny}
\affil{ Nicolaus Copernicus Astronomical Center, Bartycka 18,
            00-716 Warsaw, Poland}
\email{agnes@camk.edu.pl; bcz@camk.edu.pl} 

\author{Aneta Siemiginowska}
\affil{Harvard Smithsonian Center for Astrophysics, 60 Garden
Street, MA02138, Cambridge, USA}
\email{aneta@head-cfa.harvard.edu}

\begin{abstract}

The time dependent evolution of the accretion disk around black hole is 
computed. The classical description of the $\alpha$-viscosity is adopted so 
the evolution is driven by the instability operating in the innermost 
radiation-pressure dominated part of the accretion disk. We assume that the 
optically thick disk always extends down to the marginally stable orbit so it 
is never evacuated completely. We include the effect of the advection, 
coronal dissipation and vertical outflow. We show that the presence of 
the corona 
and/or the outflow reduce the amplitude of the outburst. If only about half of 
the energy is dissipated in the disk (with the other half dissipated in the 
corona and carried away by the outflow) the outburst amplitude and duration 
are consistent with observations of the microquasar 
GRS 1915+105. Viscous evolution 
explains in a natural way the lack of direct transitions from the state C 
to the state B in color-color diagram of this source. Further reduction of 
the fraction of energy dissipated in the optically thick disk switches off 
the outbursts which may explain why they are not seen in all high 
accretion rate sources being in the Very High State.

\end{abstract}


\section{Introduction}

Accreting black holes are strongly variable sources in all ranges of 
timescales. Short time-scale variability is usually described through the
Fourier analysis and the state of the source can be identified by a
characteristic shape of the power spectra of galactic black holes (GBH).
 Physical interpretation
of this variability is not known although some hypotheses were formulated.
Since the power spectra of X-ray emission for sources in the Low (hard)
and High (soft) state are clearly different the underlying mechanism may 
also be different in both cases. Furthermore, the Very High State is also 
characterized by
distinctively different variability and spectral properties so yet another
mechanism may be responsible for variations in that spectral state.
The nature of this particular variability is the subject of our paper.

Belloni et al. (2000) performed a detailed observational analysis of
the variability patterns in the microquasar GRS 1915+105.  It was
previously found (Belloni et al. 1997a, 1997b), that during the high
luminosity states the spectrum of this source is dominated by a
thermal, disk-like component and the variability was modeled
phenomenologically by periodic disappearance of the inner accretion
disk (Feroci et al. 1999).  Nayakshin, Rappaport \& Melia (2000) and
Janiuk, Czerny \& Siemiginowska (2000) suggested recently, that the 
variability of the
microquasar GRS 1915+105 in its Very High State displays features
characteristic for the limit cycle behaviour well known in CV but with
the timescales orders of magnitude shorter.

The limit cycle behaviour
of CV is caused by the disk thermal and viscous instability in the region of
partial ionization of the hydrogen (Smak 1984, Meyer
\& Meyer-Hofmeister 1984) and was studied in the context of time
evolution of these objects (Cannizzo 1993, Hameury et al. 1998).  The
same instability is also responsible for the eruptions of many X-ray
transients, including GBH, and its characteristic timescales are
days/years (Cannizzo et al. 1995).  As the accretion disks around
supermassive black holes are expected to have similar properties, also
for these objects the analysis of the thermal and viscous instability was
performed (Lin \& Shields 1986, Mineshige \& Shields 1990). The
evolution of the accretion disk in AGN driven by the partial
ionization of the hydrogen was computed e.g. in Siemiginowska et
al. (1996).

However, the standard model of
accretion disk by Shakura \& Sunyaev (1973) predicts also the existence of
additional instability due to the radiation pressure, with timescales orders
of magnitude shorter, which is therefore a
good candidate to explain the behaviour of the microquasar, as well as
the short term variability in AGNs. This radiation pressure
instability was first noticed in Pringle, Rees \&Pacholczyk (1973) and
studied in Lightman \& Eardley (1974) and Shakura \& Sunyaev (1976).

In order to obtain a limit cycle behaviour in the range of radiation
pressure instability, some stabilizing mechanism must be taken into
account. Abramowicz et al. (1988) found that radial advection has
stabilizing effect on the disk at high accretion rates. This was
adopted by Honma et al. (1991) in the disk evolution
computations. Thermal stability of the transonic accretion disks was
studied in Szuszkiewicz \& Miller (1997) and
the limit cycle behaviour was confirmed by calculations of
Szuszkiewicz \& Miller (1998).

Detailed analysis of the stability properties of the disk models in
comparison with the observational data leads to the conclusion that in
order to obtain better qualitative agreement with the observations,
certain modifications of the theoretical models are necessary.
Szuszkiewicz (1990) studied slim disk models with 
viscosity prescriptions different from standard $\alpha$ viscosity of
Shakura \& Sunyaev (1973) in order to obtain appropriate outburst amplitudes. Also
 other  modified viscosity laws were introduced 
(e.g. Milsom et al. 1994, Taam et al. 1997).
Moreover, several authors  postulated
that $\sim 90\%$ of the energy generated in the disk is radiated away
by the disk corona (Nayakshin et al. 2000) or that energy carried
by jet has to be included in the model (Janiuk, Czerny \& Siemiginowska 2000). 
However, while the last modification seems to be well supported by the
observational data, the dominating role of the hot corona is clearly
in contradiction with
the observed spectral shape of the sources in the Very High
State.

In the present paper we propose the time dependent model of an
accretion disk and we calculate the disk evolution based on radiation
pressure instability. We study in detail the model properties and its
sensitivity to the adopted parameters.  
We consider the role of advection in the energy budget and we take
into account the possible effects of an outflow and a corona,
 which are
consistent with our current understanding of the accretion flow at 
high luminosity to the Eddington luminosity ratio. Finally, we compare
the  resulting lightcurves and spectra with the observations.   

\section{Radiation pressure instability}

In this section we analyze the local instability of an accretion disk
caused by the radiation pressure domination.
The simplest approach to the study of the disk stability is based on the
vertically averaged description of the disk structure (Shakura \& Sunyaev 1973,
Shakura \& Sunyaev 1976). 
The vertically averaged model offers an easy qualitative insight
into the disk properties. However, this approach is neither accurate nor 
unique. Various
approaches to the averaging were used in the literature and they differ 
in detailed predictions of the disk stability.

Paczy\' nski \& Bisnovatyi-Kogan (1981) introduced the
politropic approximation to the disk vertical structure which resulted in
the presence of additional numerical factors, ranging from 0.5
to 6.0, in the averaged equations. The same values were later adopted by 
Abramowicz et al. (1988).
Honma et al. (1991) adopted different coefficients ranging from 1.0 to
16.0 and in Nayakshin, Rappaport \&
Melia (2000) the coefficients were in the range from 0.5 to 2.0. The 
stability of
the disk with advection
was also studied within the frame of the vertically averaged quantities
(Abramowicz et al. 1988, Szuszkiewicz 1990, Szuszkiewicz, Malkan \& Abramowicz
1996). Here we stress that the values of the accretion rate corresponding 
to the transition
from the stable branches to the intermediate unstable branch may well be
affected by this approach.

Recent simulations of the local structure of radiation pressure dominated disks
performed by Agol et al. (2001), Turner \& Stone (2001) and Turner, Stone
\& Sano (2002) agreed with the analytical models. They confirmed that the 
radiation pressure dominated disks are unstable, although details of the 
viscous heating and convection may affect the accretion rate at which the 
instability occurs. Interestingly, these first simulations also show that 
the result of the instability could lead in some cases to the disk collapsing 
to a new gas pressure dominated equilibrium. The long term global evolution 
effects of the radiation pressure dominated disks have not been calculated 
so far in the 2D simulations.
   
Since the full 2D approach to the study of global long timescale disk 
evolution is very complex, we adopt the 1D+1D approach.   
Therefore in subsection 2.1
we study the full vertical structure of the accretion disk with standard 
prescription for
the viscosity and advection.
Due to this approach in subsection 2.2
we are able to determine uniquely the appropriate
numerical coefficients, which are subsequently used in the vertically averaged
disk model. In subsection 2.3 we include in the model the presence of a hot
corona above the disk and a possibility of a vertical outflow powered by
accretion. In subsection 2.4 we describe the shape of the stability
curve  resulting from the averaged equations and we discuss the role of this
 stability curve and of the model parameters in the time evolution of the 
accretion disk.

\subsection{Vertical structure of a stationary accretion disk}
\label{sec:ver}

The disk model at a given radius $r$ is parameterized by the mass of the black
hole $M$, accretion rate $\dot M$, and the viscosity parameter $\alpha$.
In this work we use a standard viscosity prescription, based on the assumption
 that the viscous stress tensor is proportional to the total pressure $P$:
\begin{equation}
\tau_{\rm r\varphi}=-\alpha P.
\end{equation}

While presenting the results, we frequently use the dimensionless accretion 
rate $\dot m$, given by 
\begin{equation}
\dot m \equiv{\dot M\over \dot M_{\rm Edd}},
\end{equation}
where $\dot M_{\rm Edd}$ is the critical (Eddington) accretion rate: 
\begin{equation}
\dot M_{\rm Edd} = {L_{\rm Edd} \over c^2 \eta}={4\pi G M m_H \over \sigma_T c \eta}.
\label{eq:edd}
\end{equation}
The efficiency of accretion is $\eta= 1/16$, as it results from the 
pseudo-Newtonian approximation to the disk accretion 
(Paczy\' nski \& Wiita 1980).  

The total energy flux $F_{\rm tot}$ dissipated within the disk at a radius $r$ is 
determined by the global 
parameters:

\begin{equation}
F_{\rm tot} = {3 G M \dot M \over 8 \pi r^3} f(r),
\label{eq:ftot}
\end{equation}
where $f(r)$ represents the boundary condition at the mar\-gi\-nal\-ly stable 
orbit:
\begin{equation}
f(r)=1-({r_{\rm ms} \over r})^{3/2} {r-r_{\rm g} \over r_{\rm ms}-r_{\rm g}}
\label{eq:fr}
\end{equation}
in the pseudo-Newtonian approximation.

The disk vertical structure is described by a set of basic equations:
the equation of 
viscous energy dissipation,
\begin{equation}
{d F \over d z}=\alpha P (-{d \Omega_K \over d r}),
\label{eq:prod}
\end{equation}
the hydrostatic equilibrium,
\begin{equation}
{1 \over \rho} {dP \over dz} = - \Omega_K^2 z,
\label{eq:hydro} 
\end{equation}
where the total pressure is the sum of the gas and radiation pressure:
\begin{equation}
P = P_{\rm gas} + P_{\rm rad},
\end{equation} 
and equation of energy transfer,
\begin{equation}
{d T \over d z}=-{3 \kappa \rho \over 4 a c T^{3}} F_{\rm l}.
\label{eq:trans}
\end{equation}
 The latter takes into account the 
presence of convection which carries non-negligible fraction of energy
(see e.g. Shakura, Sunyaev \& 
Zilitinkevich 1978, D\" orer et al. 1996):
\begin{eqnarray}
\lefteqn{~~~~~~~ F_{\rm l}=F_{\rm rad}
 ~~~~~  \nabla_{\rm rad}\le\nabla_{\rm ad} {}}
\nonumber \\
 & & {} F_{\rm l}= F_{\rm rad} + F_{\rm conv} ~~~~~ 
\nabla_{\rm rad}>\nabla_{\rm ad} {}
\end{eqnarray}
Here $F_l$ is the energy flux transported locally in the direction 
perpendicular to the equatorial plane, carried either by radiation or by 
convection. It differs from the flux $F$ due to the presence of radial
advection. We introduce the advection term into our set of equations assuming
that the fraction of energy radially advected does not depend on the distance 
from
the equatorial plane.
Such an approach seems to be justified, since the dependence of the radial 
velocity on the distance from
the equatorial plane is uncertain and it does not influence other equations. 
Therefore we calculate the required
entropy gradients at the equatorial plane, denoting the appropriate 
thermodynamical quantities with an index $e$. 
The local energy flux is then given by
\begin{equation}
F_l=F_{\rm tot}(1-f_{\rm adv})
\end{equation}
and the advected fraction $f_{\rm adv}$ is determined from the global 
ratio of the total
advected flux to the total viscously generated flux (e.g. Paczy\' nski \&
Bisnovatyi-Kogan 1981, Muchotrzeb \&
Paczy\' nski 1982, Abramowicz et al. 1988):
\begin{equation}
f_{\rm adv}= {F_{\rm adv} \over F_{\rm tot}}=
- {2 r P_e q_{\rm adv} \over 3 \rho_eGM f(r)},
\end{equation}
where
\begin{equation}
q_{\rm adv}=(12-10.5 \beta_e) {d ln T_e \over d ln r} - 
(4-3 \beta_e){d ln \rho_e \over d ln r}.
\label{eq:qadv}
\end{equation}
Here $\beta$ is the gas pressure to the total pressure ratio
\begin{equation}
\beta = {P_{\rm gas} \over P}.
\end{equation}

Defining the advection term through the quantities at the disk equatorial 
plane is an oversimplification which reflects the fact that the approach
to the vertical structure of the disk outlined above does not include any
information about the vertical distribution of the radial velocity. The
actual flow pattern can be studied only if we allow for the departure
from the strict vertical hydrostatic equilibrium. The computed flow
pattern in Keplerian disks is rather complex since the meridional circulation
is superimposed on the mean accretion flow (e.g. Siemiginowska 1988). 
Therefore,
there is some uncertainty involved at this stage even in the case of
disk vertical structure included.

The frequency-averaged opacity $\kappa$ (Rosseland mean) used in
vertical structure computations
includes the electron scattering as
well as the required free-free and bound-free transitions. The tables are
from Alexander, Johnson \& Rypma (1983) 
for log $T <3.8$, from Seaton et al. (1994) for  log $T >4.0$ and the value
of opacity is 
interpolated between these two tables for intermediate values of the
temperature, as in R\' o\. za\'nska et al. (1999).  
The numerical code which computes the disk vertical structure was
developed from the version  from Pojma\' nski (1986)
and subsequently modified by R\' o\. za\' nska et al. (1999).

If the value of the $f_{\rm adv}$ is specified somehow in advance the integration 
of the disk vertical structure proceeds in a usual way. The boundary
conditions to the set of differential equations are 
specified in the equatorial plane $z=0$ and at the disk surface
located at a priori unknown height $z=H$:
\begin{eqnarray}
F(0)=0, ~~F(H) = F_{\rm tot},
\nonumber \\
\sigma T^{4} (H)=0.5 F_{\rm tot}(1-f_{\rm adv}). 
\end{eqnarray}

The presence of radial derivatives means that the problem in general should be 
solved as a global one, after defining the appropriate outer and inner radial 
boundary conditions. However, here for simplicity we adopt the local approach
to advection developed and successfully used by  Chen (1995). 
The numerical value of $q_{\rm adv}$, given by Equation \ref{eq:qadv},
may be specified at each radius by iterations at the two neighbouring radii.
This advection term does not depend much on the accretion rate and may well be assumed constant on  the thermal equilibrium curve for a given radius.
Such a method is 
correct for disks which are Keplerian and not too close to the marginally
stable orbit, where the conditions close to the sonic point are essential
(see also Yuan 1999).

\subsection{Vertically averaged stationary disk model}
\label{sec:aver}

In the vertically averaged disk structure the 
differential equations (\ref{eq:prod}) - (\ref{eq:trans})
may be replaced by the quantities defined as  mean values or values determined 
at the equatorial plane.
The numerical coefficients $C_{\rm i}$ are uniquely  determined here from the
 vertical structure model:

\begin{equation}
C_{\rm 1} = {\int_{0}^{H} P dz \over P_{\rm e} H}
\end{equation}

\begin{equation}
{1 \over C_{\rm 2}} = {\int_{0}^{H}\rho F dz \over \rho_{\rm e} F_{\rm rad} H}
\end{equation}

\begin{equation}
C_{3} = {\int_{0}^{H} \rho z dz \over \rho_{\rm e} H^2}
\end{equation}

Therefore we can use the vertically averaged relations:

\begin{equation}
F_{\rm tot}= C_{1} {3\over 2}\alpha P \Omega_K H
\end{equation}
is the local energy flux generated through the $\alpha$ viscosity,
\begin{equation}
F_{\rm l} H=C_{2} {a c T^{4} \over 3 \kappa \rho} 
\end{equation}
is the local emitted flux, and the hydrostatic equilibrium is expressed as:
\begin{equation}
{P \over \rho} = C_{3} \Omega_K^2 H^2. 
\label{eq:hyd}
\end{equation}

From the disk vertical structure calculations we obtained numerical 
coefficients $C_{1}$,  $C_{2}$ and $C_{3}$ approximately equal to 0.4, 2.0 
and 0.17  \footnote{Our factors $C_{\rm i}$  correspond to the coefficients 
defined in Muchotrzeb \& Paczy\' nski (1982): $B_{1}=0.8$,  $B_{3}=5$ and 
$B_{4}=6$, while these authors use $B_{1}=0.5$,  $B_{3}=B_{4}=6$ and Honma 
et al. (1991) use 
$B_{1}=1.0$,  $B_{3}=16$ and $B_{4}=8$.} (they only weakly depend on the 
accretion rate and viscosity).
Using these coefficients we can write for any given radius the
vertically averaged, time-independent, energy balance  equation:

\begin{equation}
C_{1} H {3\over 2}\alpha P \Omega_K =  C_{2} {1 \over H} {a c T^4 \over 3 \kappa \rho} - {1 \over 2 \pi r} \dot M T {dS \over dr}, 
\label{eq:stat}
\end{equation}
where the left hand side refers to the viscous energy dissipation
 and the right hand side describes energy losses due to diffusion of 
radiation and advection. The entropy gradient is equal to:
\begin{equation}
T {dS \over dr} r = {P \over \rho} q_{\rm adv}
\label{eq:entrop}
\end{equation}
where $q_{\rm adv}$ is described by the Equation (\ref{eq:qadv}) and is approximately constant.
The energy balance Equation (\ref{eq:stat}) together with the averaged
hydrostatic equilibrium Equation (\ref{eq:hyd}) allow us to compute the
stability curve on the surface density versus temperature ($\Sigma - T$) plane.
These stationary solutions are presented in Section \ref{sec:local}.

\subsection{Stationary disk with a corona and vertical outflow}
\label{sec:stat}

Detection of the power law tail in the hard X-ray spectra
of black hole X-ray binaries and AGNs (Mushotzky, Done \& Pounds 1993, 
Tanaka \& Lewin 1995) indicates, that somewhat
hotter medium is present in the vicinity of a standard accretion
disk. Here we assume, that this hot gas forms a corona above the disk,
which extends down to the marginally stable orbit and dissipates a
constant fraction $f_{\rm cor}$ of the total accretion energy. 
This energy does not contribute to the disk black body emission, but
is stored in the corona and radiated in the hard X-rays.

For a number of objects the presence of a collimated vertical
outflow is confirmed (e.g. Mirabel \& Rodriguez 1994 for GRS
1915+105). We also include in our model such a possibility, 
however we do not adopt any collimation mechanism. The
plasma ejection is powered by the energy released in the innermost
disk parts. Therefore we assume that another fraction of a total
energy is converted into the  wind and is not seen in the disk multicolor
black body emission.  We parameterize this fraction using the
luminosity to the Eddington luminosity ratio, so that for accretion
near the Eddington limit we should expect the outflow to be the strongest.
 Unlike in the  ADAF models, where a powerful wind may take away a 
considerable amount of mass and angular momentum (Blandford \& Begelman 1999),
 we assume, that in our disk the carried mass and angular momentum
 may be negligible (see Discussion). Therefore the presence of a wind is 
reflected only in the energy equation.
   
The energy balance Equation (\ref{eq:stat}) is modified by the outflow and
the corona.  In principle, in the time {\it independent} model, we
could put the outflowing and coronal fractions either in the cooling or in the
heating term with no change in the final energy balance. 
However, in the physical case both the outflow and the corona play their
role in cooling of the accretion flow, so in general the emitted flux
will have the form:


\begin{equation}
F^{-} =C_{\rm 2} {1 \over H}{a c T^{4} \over 3 \kappa \rho} [(1-f_{\rm cor})(1-f_{\rm out})]^{-1}, 
\end{equation}

where
\begin{equation}
f_{\rm out} = 1 - {1 \over 1+A\dot m^{2}}
\label{eq:fjet}
\end{equation}
we take after Nayakshin et
al. (2000). Our  model parameters are $A$ and $f_{\rm cor}$, which we assume to
be constants, ranging from 0 to 1.
Below we describe the role of these parameters in the resulting stationary
disk solutions.

\subsection{The local stability curve}
\label{sec:local}

When the energy balance given by Equation \ref{eq:stat} is satisfied,
the stationary local disk solutions form a characteristic {\bf S}
curve on the $\Sigma - T$ plane (alternatively on the vertical axis 
we may plot accretion
rate $\dot M$ instead of the temperature). In this section we consider the
curves calculated for the $10 M_{\rm \odot}$ black hole.  In the case
of a supermassive black hole the disk temperature is $\sim 2$ orders
of magnitude lower and the surface densities may reach $\sim 10$ times
higher values for a given $R/R_{Schw}$. It results in shifting the entire 
stability curve down and right on the $\Sigma - T$ diagram.

The accretion disk is thermally and viscously unstable when
${\partial \dot M \over \partial \Sigma}<0$ (Lightman \& Eardley 1974, 
Meyer \& Meyer-Hofmeister 1981). This is the case  for the accretion rates
in the range $(\dot m_{1}, \dot m_{2})$, where the 
{\bf S} curve has a negative slope.
The lower branch of the curve represents stable solutions, resulting from the
 gas pressure domination. The intermediate unstable branch appears due
to radiation pressure domination, while the upper hot branch is
stabilized by advection (Abramowicz et al. 1988).  The 
 upper and lower transition points $\dot m_{1}$ and $\dot m_{2}$ 
depend on the global model parameters
(i.e. $M$ and $\alpha$) as well as on the local distance $r$. The
dependence on the global parameters is relatively weak, but there is a
strong dependence on the radial location $r$: a change of $r$ results
mostly in a strong shift of the entire curve up or down on the
diagram. It translates into strong dependence of the extension of the
radiation-pressure dominated disk region on the accretion rate
(e.g. Hure 1998, Janiuk \& Czerny 2000).

\begin{figure}
\plotone{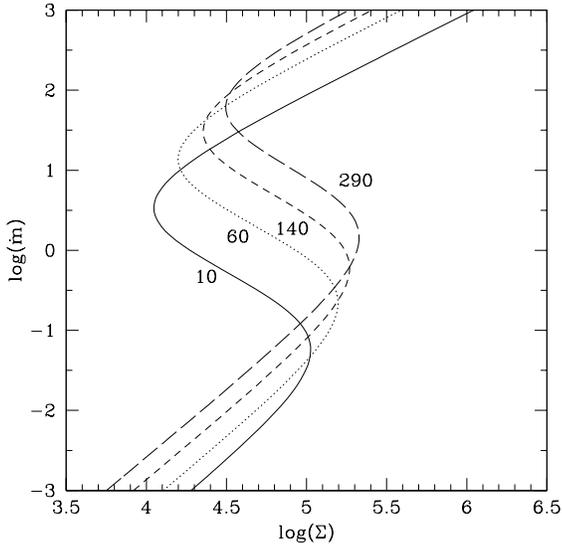}
\caption{The stability curves calculated for $M=10 M_{\odot}$,  and 4
different radii : $R=10 R_{\rm Schw}$ (solid curve), $R=60 R_{\rm Schw}$
(dotted curve), $R=140 R_{\rm Schw}$ (short-dashed curve) and $R=290
R_{\rm Schw}$ (long-dashed curve). The viscosity parameter $\alpha =
10^{-2}$.  The effects of outflow and corona are neglected.  Note, that on
the vertical axis we plot local dimensionless accretion rate,
 $\dot m= \dot M/ \dot M_{Edd}$, instead
of the disk temperature.
\label{fig:srad}}
\end{figure}

In Figure \ref{fig:srad} we plot the solutions for several disk radii.
If the external accretion rate does not exceed a critical value, the
whole disk remains in the cold, stable regime and the matter is
transported inwards with constant velocity. However, for higher
external accretion rates the inner disk parts may find themselves in
an unstable mode. Then the disk undergoes oscillations, heating up and
cooling down in a cycle. This may cause the observed luminosity
outbursts or flickering.  The range of accretion rates for which the
disk is locally unstable varies across the disk. It is wider at small
radii and becomes narrower at larger radii. This effect is even
stronger, if the full vertical structure is computed instead of the
constant $C_{\rm i}$ coefficients in the vertically averaged model. In
this case at large radii, $r > 200 R_{\rm Schw}$
 the whole intermediate branch of the {\bf S}
curve gets a positive slope and the instability vanishes.
As a result the outer
parts of the disk are stable, while the inner parts may undergo
oscillations.

The critical accretion rate, above which the disk becomes unstable, as
well as the outbursts amplitude depend on the location of the turning
points $\dot m_{1}$ and $\dot m_{2}$ on the stability curve. 
This is determined by 
the presence and the strength of the disk hot corona and/or outflow.

\begin{figure}
\plotone{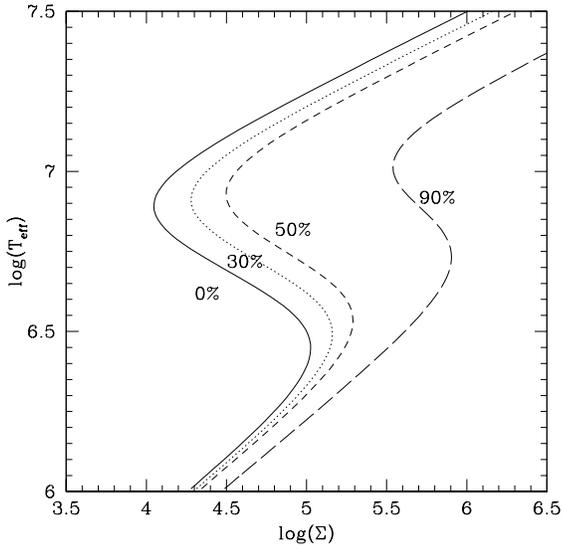}
\caption{The stability curves calculated for $M=10 M_{\odot}$, $R=10
R_{\rm Schw}$ and 4 values of the fraction of energy dissipated in the
corona: $f_{\rm cor}=0.0$ (solid curve), $f_{\rm cor}=0.3$ (dotted curve), 
$f_{\rm cor}=0.5$  (short-dashed
curve) and $f_{\rm cor}=0.9$ (long-dashed curve). The viscosity parameter 
$\alpha =
10^{-2}$. The outflow effect is neglected.
\label{fig:scor}}
\end{figure}

In Figure \ref{fig:scor} we plot the stability curves at location
$R=10 R_{\rm Schw}$, resulting from different assumptions about the
fraction of energy dissipated in the corona. When the corona is
neglected, the lower turning point corresponds to a very small
accretion rate ($\dot m_{1} \sim 0.06$) and the outburst amplitudes should
be very large. The presence of the corona increases the critical
accretion rate and reduces the size of the unstable branch, which
results in a reduction of the outbursts amplitude.

\begin{figure}
\plotone{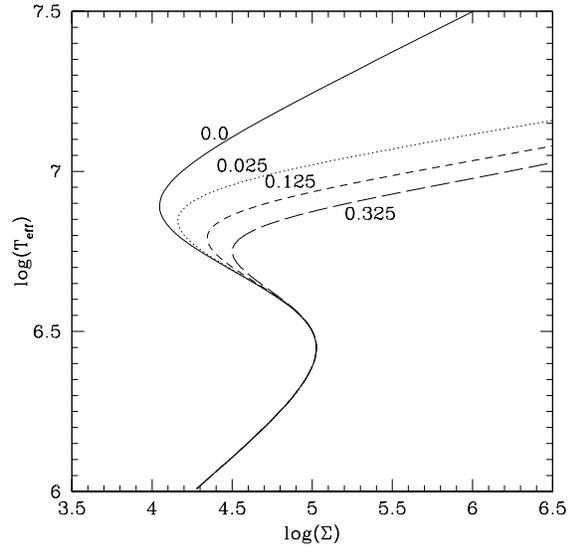}
\caption{The stability curves calculated for $M=10 M_{\odot}$, $R=10
R_{\rm Schw}$ and 4 values of the outflow parameter A (see Equation \ref {eq:fjet}): 0.0
(solid curve), 0.025 (dotted curve), 0.125  (short-dashed
curve) and 0.325 (long-dashed curve). The viscosity parameter $\alpha =
10^{-2}$.
The  effect of the corona is neglected.
\label{fig:sdzet}}
\end{figure}

In Figure \ref{fig:sdzet} we plot the solutions obtained for different
strengths of the outflow.  The strong outflow suppresses the instability at
high temperatures and reduces the amplitude of an outburst. 
However, it does not
affect the cold stable branch and has no influence on the lower
turning point on the stability curve.

\begin{figure}
\plotone{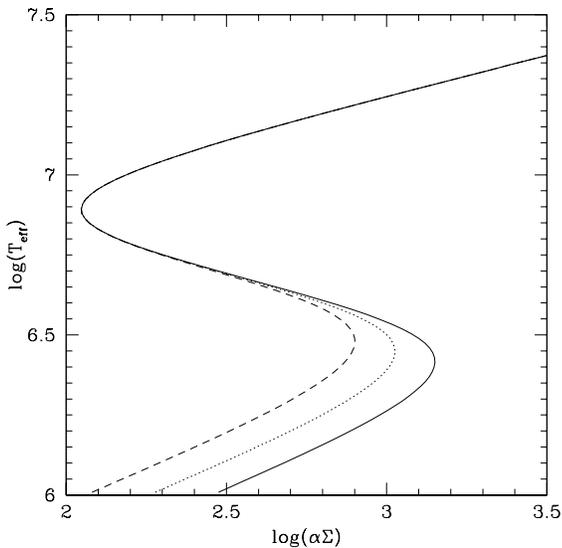}
\caption{The stability curves calculated for $M=10 M_{\odot}$, $R=10
R_{\rm Schw}$ and 3 values of viscosity parameter: $\alpha=10^{-1}$ (solid
curve), $\alpha=10^{-2}$ (dotted curve), and $\alpha=10^{-3}$ (dashed
curve). The corona and outflow effects are neglected.
\label{fig:salpha}}
\end{figure}

The evolution time scale and duration of an outburst is determined by
the viscosity. Here we use the constant viscosity parameter $\alpha$.
In Figure \ref{fig:salpha} we plot the dependence of the disk
solutions on this parameter. The stability curve is shifted almost
horizontally with $\alpha$. Therefore the viscosity parameter does not 
influence outburst amplitudes, while its value may make them last shorter 
or longer.

\section{Time dependent disk}

In this section we describe the time dependent disk model.
Here we treat the disk radial structure independently from the
vertical one, which is included in our calculations via the
coefficients $C_{\rm i}$ in the averaged equations. The crucial parameter,
which links the radial and vertical disk structures, is the viscosity
parameter $\alpha$. The viscosity mechanism is responsible for both
radial transfer of the material inwards and for the local dissipation
of energy, which in turn is either radiated vertically or
advected. Once $\alpha$ is specified, we adopt the vertically averaged
equations in the time evolution of the disk.

\subsection{The basic equations}

The energy balance expressed by the Equation (\ref{eq:stat}) gives the 
stationary solution in the $\Sigma - T$ plane. In order to construct the
 time dependent model we solve the two basic equations for $\Sigma(r,t)$ and 
$T(r,t)$, following the evolution of the accretion disk.
The mass conservation equation:

\begin{equation}
{\partial \Sigma \over \partial t}={1\over 2\pi r} {\partial \dot M \over \partial r},
\end{equation}
where
\begin{equation}
\dot M=-2\pi r \Sigma v_{\rm r},
\end{equation}
is combined with the angular momentum conservation equation:
\begin{equation}
\dot M {d\over dr}\big(r^2 \Omega\big)=-{\partial \over \partial r}\big(2\pi r^2 T_{\rm r\varphi}\big).
\end{equation}
Here $\Omega$ is the Keplerian velocity and the $r\varphi$ component of the 
stress tensor is equal to $T_{\rm r\varphi}=\alpha P H=(3/2)\Omega\nu\Sigma$.
$P$ is the total pressure, $H$ is the disk height, $\nu$ denotes the 
kinematic viscosity and $v_{\rm r}$ is the radial velocity.  
From these equations we obtain the equation of the surface density evolution:
\begin{equation}
{\partial \Sigma \over \partial t}={1\over r}{\partial \over \partial r}\Big(3 r^{1/2} {\partial \over \partial r} \big(r^{1/2} \nu\Sigma\big)\Big).
\label{eq:evolsig}
\end{equation} 
The disk temperature evolves due to the departure from the thermal energy 
conservation:
\begin{equation}
\Sigma T {d S\over d t}=Q^{+} - Q^{-}.
\label{eq:temper}
\end{equation}
Here we adopt viscous heating
\begin{equation}
Q^{+}=C_{1}(3/2)\Omega \alpha P H
\end{equation}
and radiative cooling
\begin{equation}
Q^{-}=C_{2}[(1-f_{\rm cor})(1-f_{\rm out})]^{-1}(4\sigma/3\kappa)(T^{4}/\Sigma).
\end{equation}
The coefficients $C_{\rm 1}$ and  $C_{\rm 2}$ were defined in Section 
\ref{sec:aver}. 
Combining Equation
(\ref{eq:temper}) with time-dependent forms of Equations
(\ref{eq:entrop}) and (\ref{eq:qadv}) we obtain the energy equation:

\begin{eqnarray}
{\partial \ln T \over \partial t} + v_{\rm r} {\partial \ln T \over
\partial r} =  {}
\nonumber\\
{} = {4-3\beta \over 12-10.5\beta} \Bigg( {\partial \ln
\Sigma \over \partial t} - {\partial \ln H \over \partial t} +{}
\nonumber\\
{}+ v_{\rm r}{\partial \ln \Sigma \over \partial r} \Bigg)
 + {Q^{+} - Q^{-}\over  (12-10.5\beta) P H} {} .
\label{eq:evolt}
\end{eqnarray}
The radial velocity is given by:
\begin{equation}
v_{\rm r}=-{3\over\Sigma} r^{-1/2} {\partial \over \partial r}\big(\nu\Sigma r^{1/2}\big).
\end{equation}

The corona and the vertical outflow, parameterized by $f_{\rm cor}$ 
and $f_{\rm out}$ 
(see Section \ref{sec:stat}), play a
crucial role in our calculations. In principle, they could contribute either
to the cooling or to the heating of the disk, and therefore influence
strongly the results of the time evolution. Here we choose again the 
physical case,
e.g. corona and outflow as  the disk cooling mechanisms, consistently with our
stationary solution. In the following we assume that they add to the cooling
$Q^{-}$, while the heating term $Q^{+}$ includes only viscosity.
The corona parameter $f_{\rm cor}$ is assumed constant with time and radius, 
while $f_{\rm out}$ depends on time and radius of the disk via the local 
accretion rate $\dot m(r,t)$; in this case the constant parameter of the
evolution is $A$ (see Equation \ref{eq:fjet}).

\subsection{Numerical method}

We solve the equations (\ref{eq:evolsig}) and (\ref{eq:evolt}) via
explicit method, using the constant time step, comparable to the
thermal time scale (Wallinder 1991):

\begin{equation}
t_{\rm therm}=8.75\times10^{-5} {r^{1/2}\over\alpha}(r-1)
{M\over M_{\rm \odot}}  ~~~~[s]
\end{equation}
where r is the radius expressed in Schwarzschild radii.

The viscous time scale depends on the disk thickness:
\begin{eqnarray}
t_{\rm visc}\sim {\Sigma \over {\partial\Sigma \over \partial t}} \sim
{1\over \alpha\Omega}\Big({r\over H}\Big)^{2}
\nonumber\\
t_{\rm therm}\sim {T \over {\partial T \over \partial t}}\sim
{1\over \alpha\Omega}
\nonumber\\
{t_{\rm therm}\over t_{\rm visc}}\sim \Big({H\over r }\Big)^{2}.
\label{eq:timescale}
\end{eqnarray}
For typical values of $H/r$ in our model the viscous time scales are
 3 - 4 orders of magnitude longer than the thermal
time scales.

The equation (\ref{eq:evolsig}) is transformed into a simple diffusion 
equation due to the convenient change of variables (Smak 1984):
\begin{eqnarray}
{\partial \Xi \over \partial t}={12 \over y^{2}} {\partial^{2} \over \partial y^{2}}\big(\Xi\nu\big)
\nonumber\\
y=2 r^{1/2} ~~~~~~~~ \Xi=y\Sigma
\end{eqnarray}

On the uniform radial grid in $y$ we compute in every time step the
new values of the 6 variables: $\Xi_{\rm j}^{n}$, $T_{\rm j}^{n}$,
$P_{\rm j}^{n}$, $H_{\rm j}^{n}$, $\beta_{\rm j}^{n}$ and $\nu_{\rm j}^{n}$, where $j$
is the number of radial zone and $n$ is the number of a time
step. Initial values are taken from the stationary solution on the
lower branch of the $\Sigma - T$ curve. The boundary condition on the
inner disk radius, $3 R_{\rm Schw}$, is that $T_{\rm in} = \Sigma_{\rm in} =
0$. The outer disk radius is chosen to have a stable solution,
determined by the external, constant accretion rate $\dot m_{\rm ext}$
This assumption implies, that a certain amount of the
angular momentum is taken away at the outer radius of the computed
disk region.  The typical radial extension in our calculations was
$300 R_{\rm Schw}$ and the number of points in the radial grid was 220.

\subsection{Results}

In this Section we present the results of the time dependent calculations. 
The main parameters of the model are mass of the central black hole $M$,
external accretion rate $\dot m_{\rm ext}$,
 viscosity $\alpha$, coronal fraction
$f_{\rm cor}$ and outflow strength $A$.
Throughout this work we assumed the mass of the central object equal to 10
$M_{\odot}$, which is appropriate order of magnitude for galactic black hole
 binary systems. The viscosity was constant in the disk and in every model
we adopted $\alpha=0.01$.

The external accretion rate is the accretion rate
outside the region of the radiation pressure instability, at the outer radius,
 which in  most cases was fixed at 300 $R_{\rm Schw}$. This accretion rate 
value was chosen in the range
$\dot m_{\rm ext}= 0.19-0.71$,
so that the inner disk
parts were unstable. However, we did not take into account any
effects connected with possible ionization or gravitational instability,
which may be present further out in the disk. 
We performed our calculations for the full range of corona and outflow strengths
($f_{\rm cor}$ and $A$ ranging from 0 to 1). 

Once all the model parameters are specified, the disk starts to
evolve, and tries to reach the state specified by $\dot m_{\rm ext}$. At
every disk radius the solution is found along the lower branch of the
stability curve {\bf S}, up to the moment when the innermost radii
find themselves in the unstable mode. Then the oscillations begin, and
the disk temperature and density change periodically with time. 

 Below
we present the results of our calculations. Firstly, we discuss the 
evolution at the particular disk radius during
one cycle, e.g. starting and ending on the lower stable branch of the
{\bf S} curve. This translates into one outburst of the disk total
luminosity.  Secondly, we show, how the luminosity of the disk changes
with time, depending on the assumed corona and outflow efficiencies.
Finally, we present the influence of the radiation pressure
instability on the geometrical thickness of the accretion disk, and we
show an example of the disk spectral energy distribution.

\subsubsection{Evolution of the surface density and temperature during
one cycle}

\begin{figure}
\plotone{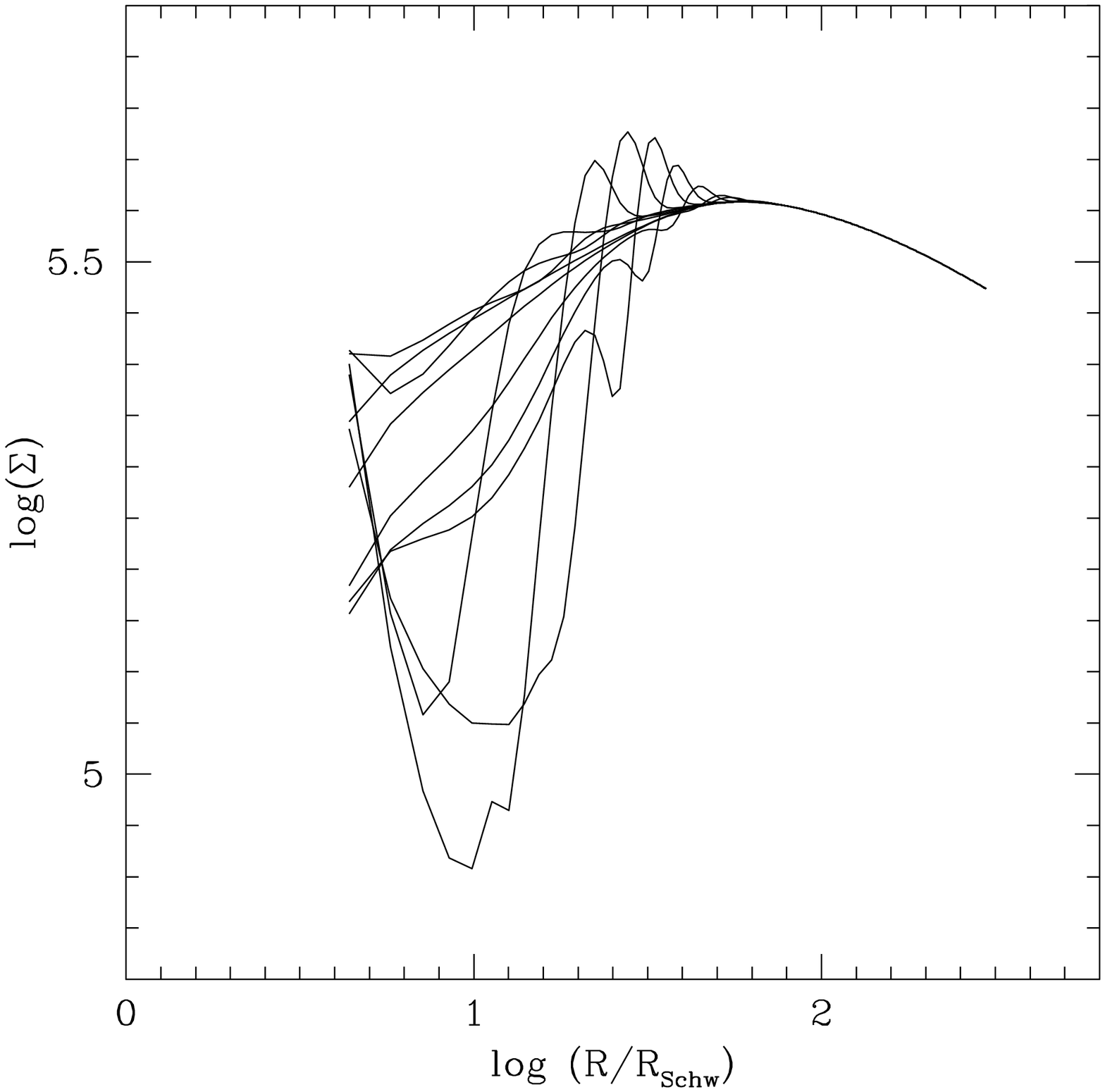}
\caption{The surface density evolution during a single outburst. Model
parameters: $M=10 M_{\odot}$, $\alpha = 10^{-2}$, $\dot m_{\rm ext} = 0.7$.
The corona share is $f_{\rm cor}=0.5$. The outflow is neglected.
\label{fig:sigma_r}}
\end{figure}

In Figure \ref{fig:sigma_r} we show the evolution of the surface
density during one cycle. Between the outbursts the radial
dependence of $\Sigma$ is smooth and has a positive slope. The radial
inflow through the accretion disk is stable and the material
accumulates slowly in the inner region. When the instability starts,
the density drops down rapidly in the innermost radii, along with the
increasing local accretion rate. The matter is transferred inwards
much faster than it is supplied from the outer disk regions and the
inner disk empties. At the outer end of the unstable region there
appears a local density peak, which shifts outwards as the instability
propagates into larger radii. The outburst finishes when there is no
more additional matter supply.  As the disk cools down, the density in
the inner part gradually rises and when the cycle is completed, the
flow returns to the initial configuration.

\begin{figure}
\plotone{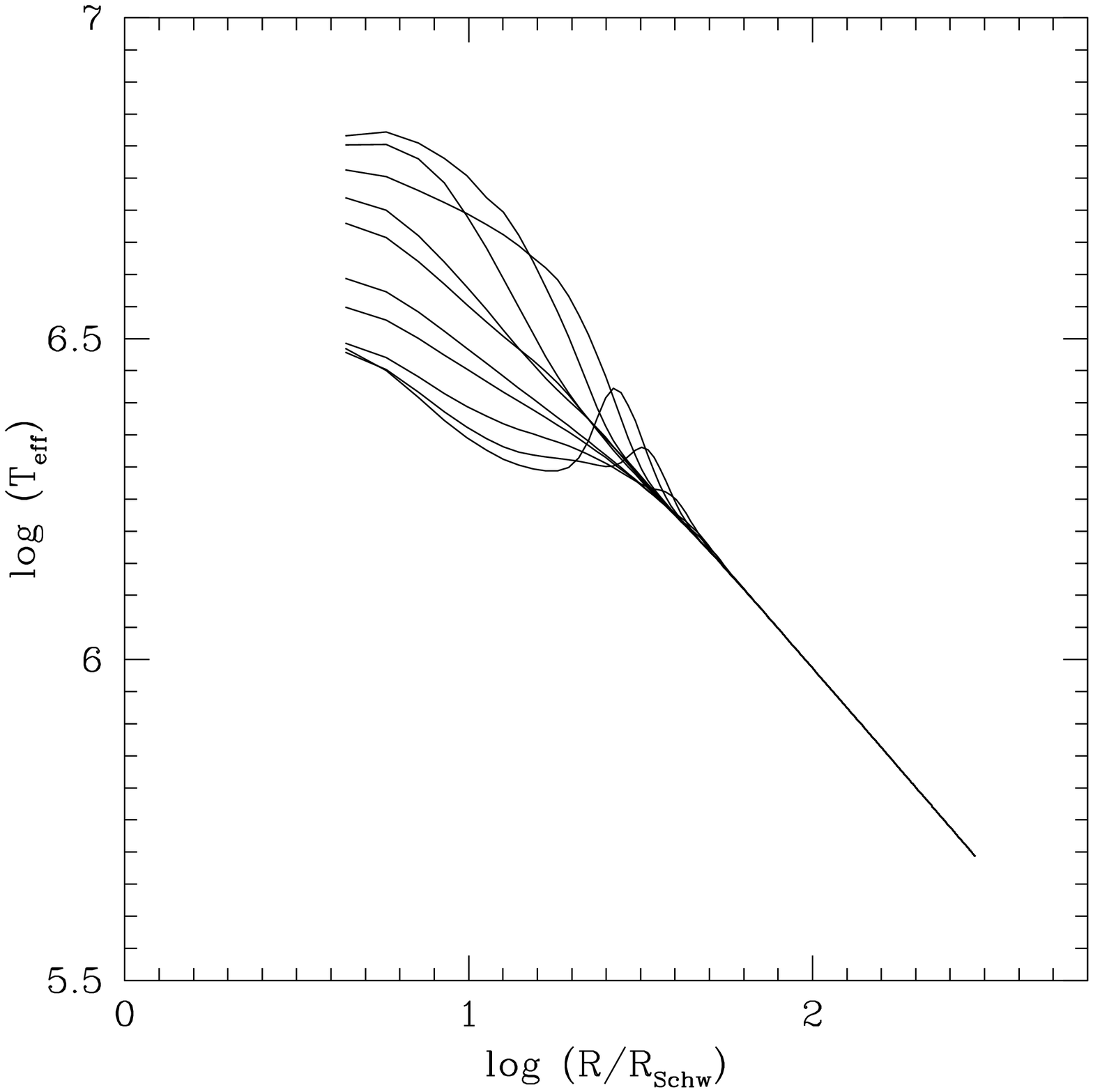}
\caption{The temperature evolution during a single outburst. Model
parameters: $M=10 M_{\odot}$, $\alpha = 10^{-2}$, $\dot m_{\rm ext} = 0.7$.
The corona share is $f_{\rm cor}=0.5$.
 The outflow is neglected.
\label{fig:t_r}}
\end{figure}

In the outburst the disk temperature rises mostly in the inner
disk regions. In Figure \ref{fig:t_r} we show the temperature
evolution during one cycle. The instability begins at the innermost
disk radius. It propagates outwards and after several time steps the
heat front reaches further regions. The temporal decrease of the
temperature in the inner part is characteristic for models with both high
external accretion rates and considerable corona share.  During this
temporal decrease the disk is still heated up in outer regions, and
this phase corresponds to the smaller ``dip'' in the lightcurve. It is
followed by the subsequent heating of the entire unstable disk region,
which is the cause of the sharp luminosity increase. At the end of the cycle
the whole disk cools down rapidly. The cooling front again starts from
the inner disk radius and propagates on the thermal time scale.  The
disk enters into the phase of relative faintness of the inner parts.
It may observationally create an impression that the inner disk has
been completely evacuated, while actually the disk always extends down
to the marginally stable orbit.

\begin{figure}
\plotone{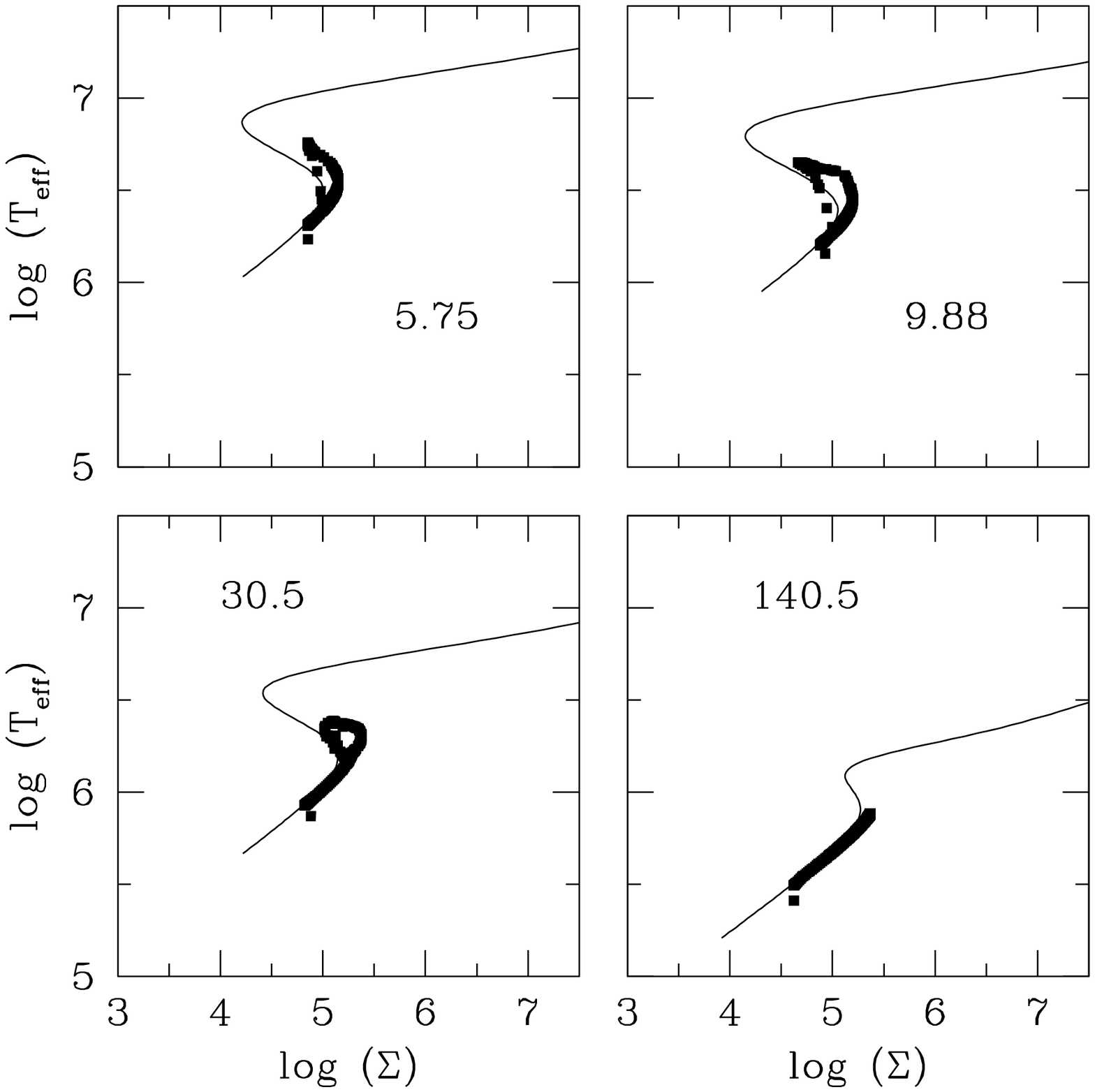}
\caption{The  evolution on the $\Sigma - T$ plane for four different
values of radius: 
$5.75 R_{\rm Schw}$, $9.88 R_{\rm Schw}$, $30.5 R_{\rm Schw}$ and $140.5
R_{\rm Schw}$.  
The  solid line marks the local stability curve and the points
represent the solutions of the time dependent disk model. We show the time
sequence from the beginning of the evolution through a few outbursts.
 Model parameters: $M=10 M_{\odot}$, 
$\dot m_{\rm ext} = 0.65$, $\alpha=0.01$.
The corona share is neglected and the outflow efficiency parameter is 
A = 0.025.  
\label{fig:sigma_t_evol}}
\end{figure}

The evolution on the $\Sigma - T$ plane depends on the local disk
radius.  The innermost parts of the disk reach the highest temperatures
and if the external accretion rate is high enough, they may even jump to
the upper, advection supported branch of the stability curve. This 
depends on other
model parameters, e.g. the strength of the corona and the outflow.
Further out the upper branch is never reached and the disk undergoes
oscillations in the intermediate unstable region. The outermost parts
of the disk always remain on the lower, cold stable branch (see Figure
\ref{fig:sigma_t_evol}).

\subsubsection{The luminosity of the disk}

The total disk luminosity is equal to $L=\int_{R_{\rm in}}^{R_{\rm
out}} 2\pi r F_{\rm rad} dr$ and it results only from the radiation of
the disk. We do not include the fraction radiated by the corona,
carried by the outflow or advected.  The coronal fraction is assumed
constant during the evolution. However, the outflow strength depends on
the luminosity to the Eddington luminosity ratio and varies: from
negligible value between the outbursts to $f_{\rm out} \sim 0.35$ during
the outburst.  Therefore, in the absence of an outflow, the resulting
outburst amplitude remains constant, being somewhat lower for a disk
with a strong corona than for a disk alone.  When the outflow is present,
the observed outburst amplitude varies during the evolution.  The
observed luminosity fraction radiated by the accretion disk  is
equal to $f_{\rm obs}=(1-f_{\rm cor})(1-f_{\rm out})$. Depending on other model
parameters, in the case of a strong outflow its luminosity may reach even
$f_{\rm out}/f_{\rm obs}=30 - 70\%$.

\begin{figure}
\plotone{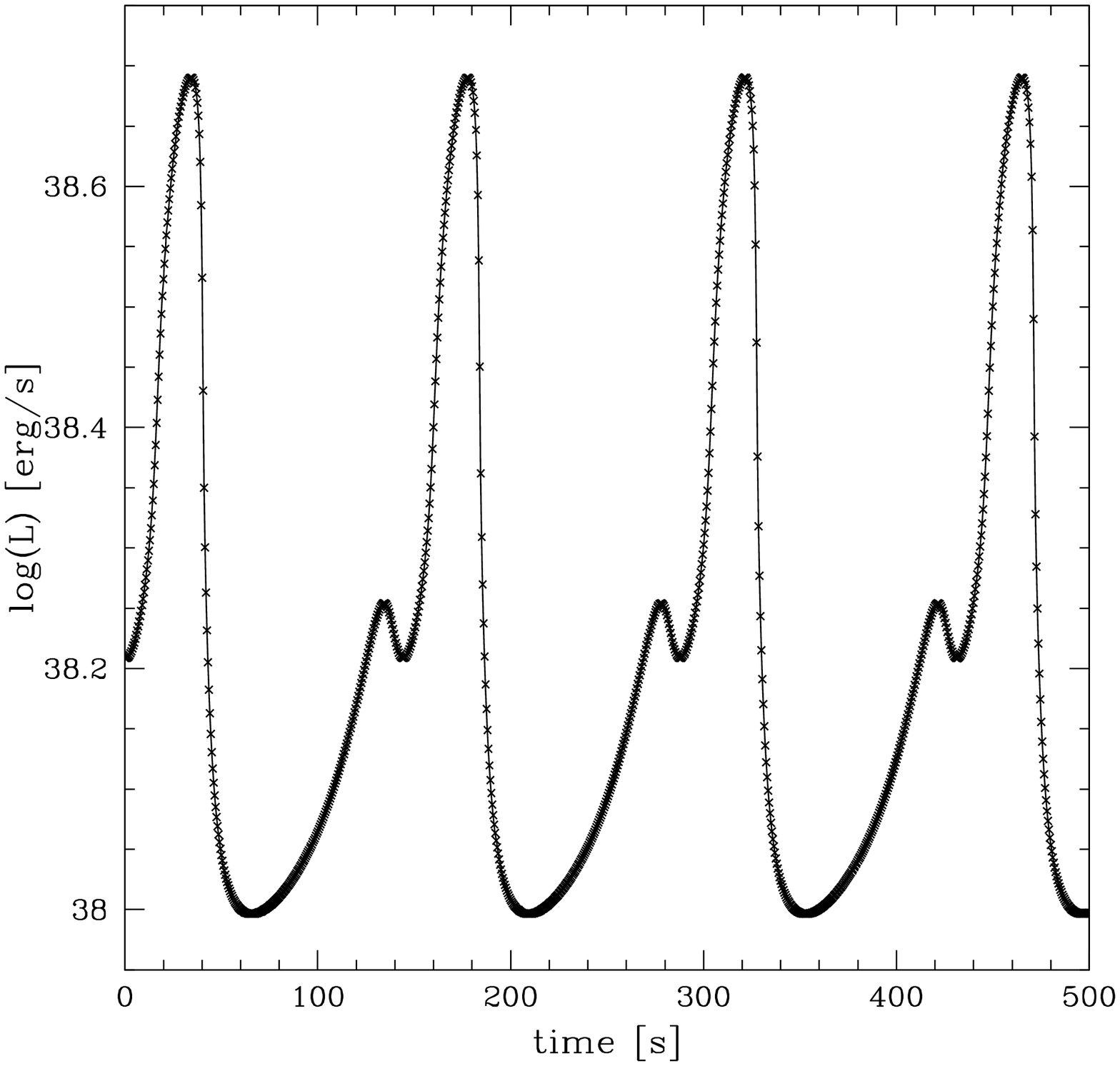}
\caption{The disk lightcurve. 
Model parameters:  $M=10 M_{\odot}$, 
$\dot m_{\rm ext} = 0.7$,
$\alpha=0.01$. The fraction of the energy
 dissipated in the corona is $f_{\rm cor}=0.5$. The outflow is neglected.
\label{fig:lum1}}
\end{figure}

\begin{figure}
\plotone{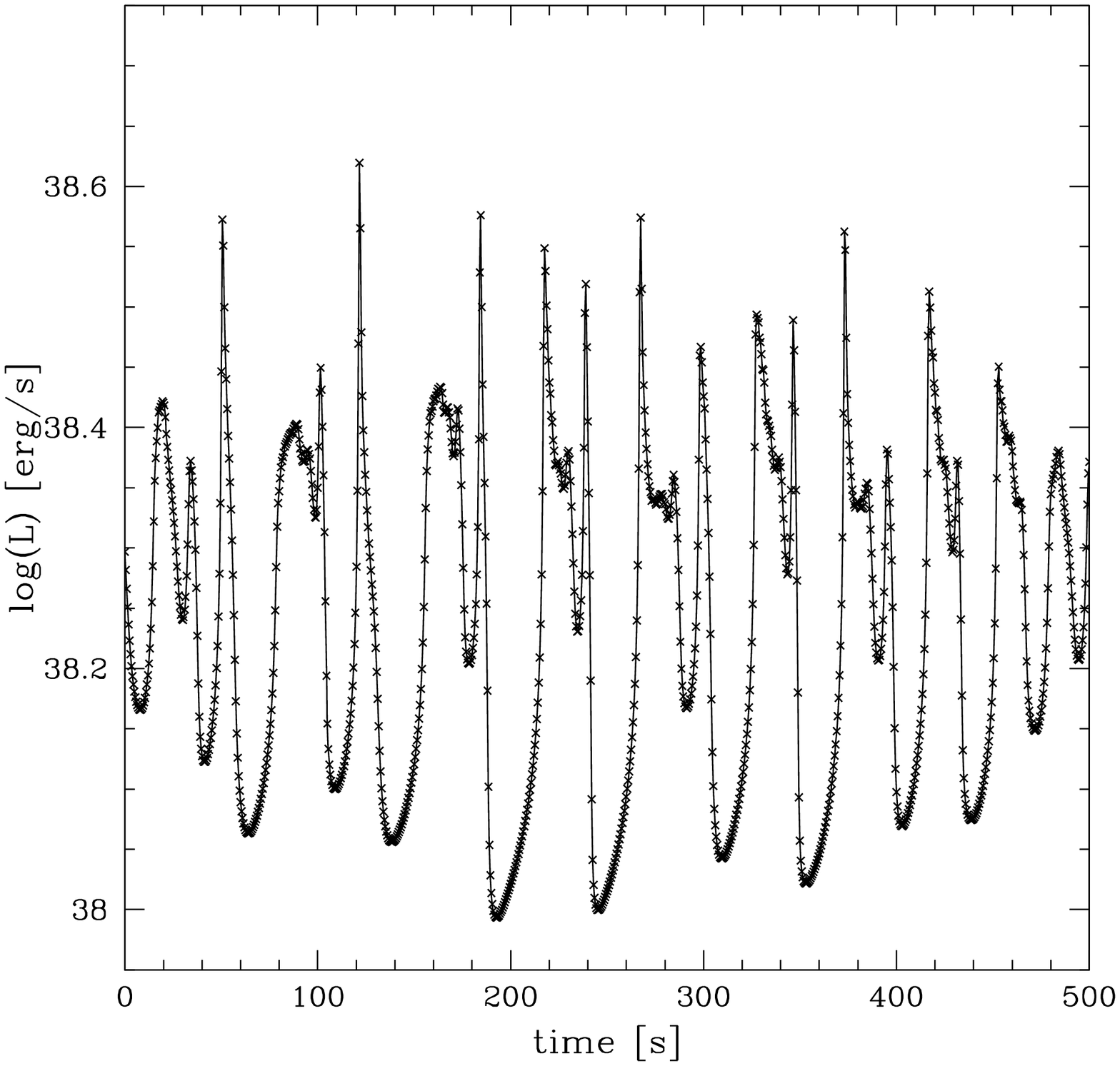}
\caption{The disk lightcurve. 
Model parameters:  $M=10 M_{\odot}$, 
$\dot m_{\rm ext} = 0.7$,
$\alpha=0.01$. The fraction of the energy 
 dissipated in the corona is $f_{\rm cor}=0.3$ and the outflow parameter is 
$A=0.075$.
\label{fig:lum2}}
\end{figure}

In Figures \ref{fig:lum1} and \ref{fig:lum2} we show the luminosity
curves for $\dot m_{\rm ext} = 0.7$.
 In the first case the outflow was not included and the
fraction of the energy dissipated in the corona was equal to $f_{\rm cor}=0.5$.
 This resulted in strong and
regular outbursts. For the model with the outflow contribution 
parameterized by $A=0.075$ the
outburst shape is quite irregular. In this case the share of the corona
has been dropped to $f_{\rm cor}=0.3$, 
which allowed us to keep the mean outburst
amplitude roughly unchanged.

\begin{figure}
\plotone{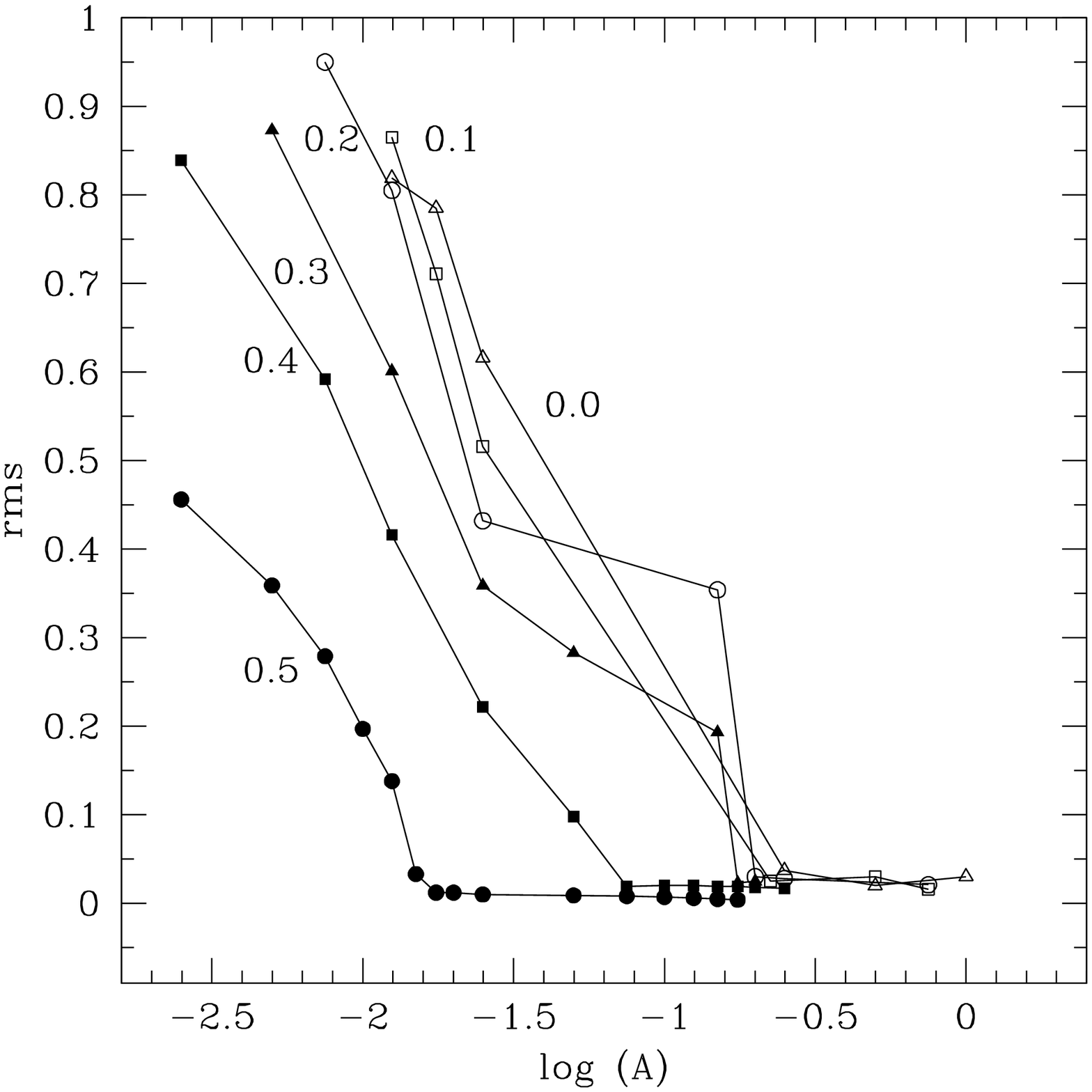}
\caption{Disk luminosity variations for different outflow and corona strengths.
The open triangles mark solutions for corona parameter $f_{\rm cor}=0$, open squares for $f_{\rm cor}=0.1$, open circles for $f_{\rm cor}=0.2$, filled triangles for 
$f_{\rm cor}=0.3$, filled squares for  $f_{\rm cor}=0.4$ and filled circles for $f_{\rm cor}=0.5$. Other model
parameters were kept constant: $M=10 M_{\odot}$, $\dot m_{\rm ext} = 0.7$,
 $\alpha=0.01$. 
\label{fig:rms}}
\end{figure}

In order to check if the luminosity variations would vanish for the
largest outflow and corona shares, we computed a grid of models 
for the full range of these two parameters. Because the outflow 
creates irregular outburst shapes and amplitudes  
we parameterized the luminosity variations using the root mean square
approximation, defined as:

\begin{equation}
rms = {\sigma \over <L>}, 
\label{eq:rms}
\end{equation}
where
\begin{equation}
\sigma^{2}={\int^{t_{2}}_{t_{1}} (L(t)-<L>)^{2} dt \over \Delta t},
\end{equation}
\begin{equation}
<L>={\int^{t_{2}}_{t_{1}} L(t) dt \over \Delta t} 
\end{equation}
The time range $(t_{1}, t_{2})$ was chosen arbitrarily and in our case it is
$(1.4, 1.41)\times 10^{5}$ sec, long enough to cover several outbursts for
each of the studied lightcurves. 
In Figure \ref{fig:rms} we plot the luminosity variations in terms of
rms for different corona and outflow strengths. Both outflow and corona
suppress the variations. Therefore the largest outbursts amplitude is
achieved for disks with no corona and a very weak outflow, while the
strong outflow and/or corona results only in
luminosity ``flickering'', which could be hardly detectable.

\subsubsection{The disk thickness}

\begin{figure}
\plotone{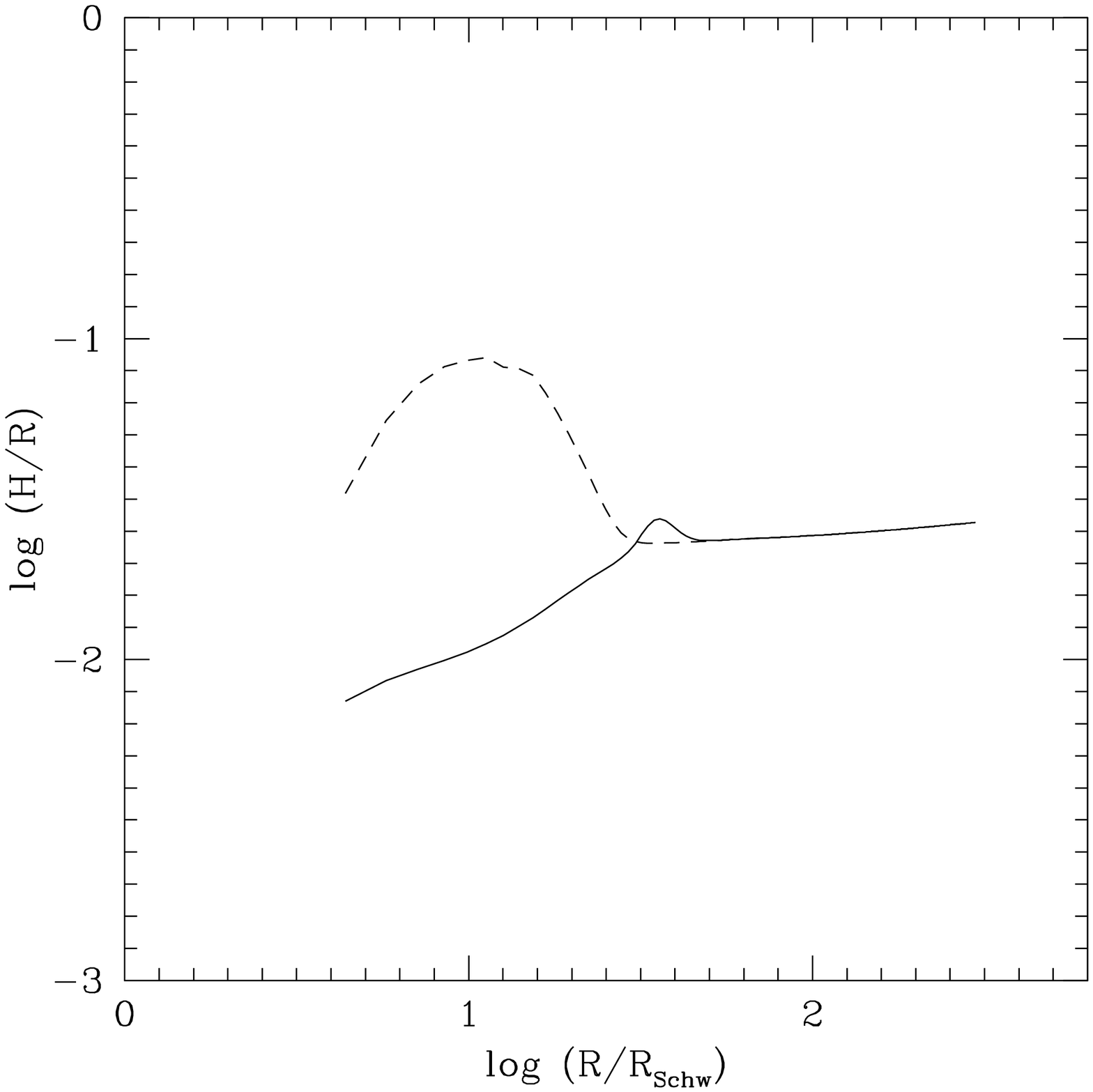}
\caption{The $H/R$ ratio during the disk
evolution in the minimum (solid line) and maximum (dashed line)
 luminosity states.
Model parameters:  $M=10 M_{\odot}$, $\dot m_{\rm ext} = 0.7$,
$\alpha=10^{-2}$. The corona share is $f_{\rm cor}=0.5$.
 The outflow is neglected.   
\label{fig:thick}}
\end{figure}

The geometrically thin disk approximation is always satisfied in our
model. However, during the outburst the inner parts of the disk become
thicker than in the cold state (see Figure \ref{fig:thick}). This is
because the temperature achieved during the outburs is controlled by the upper,
advection supported  branch of the stability  curve.

\subsubsection{Radiation spectrum}

\begin{figure}
\plotone{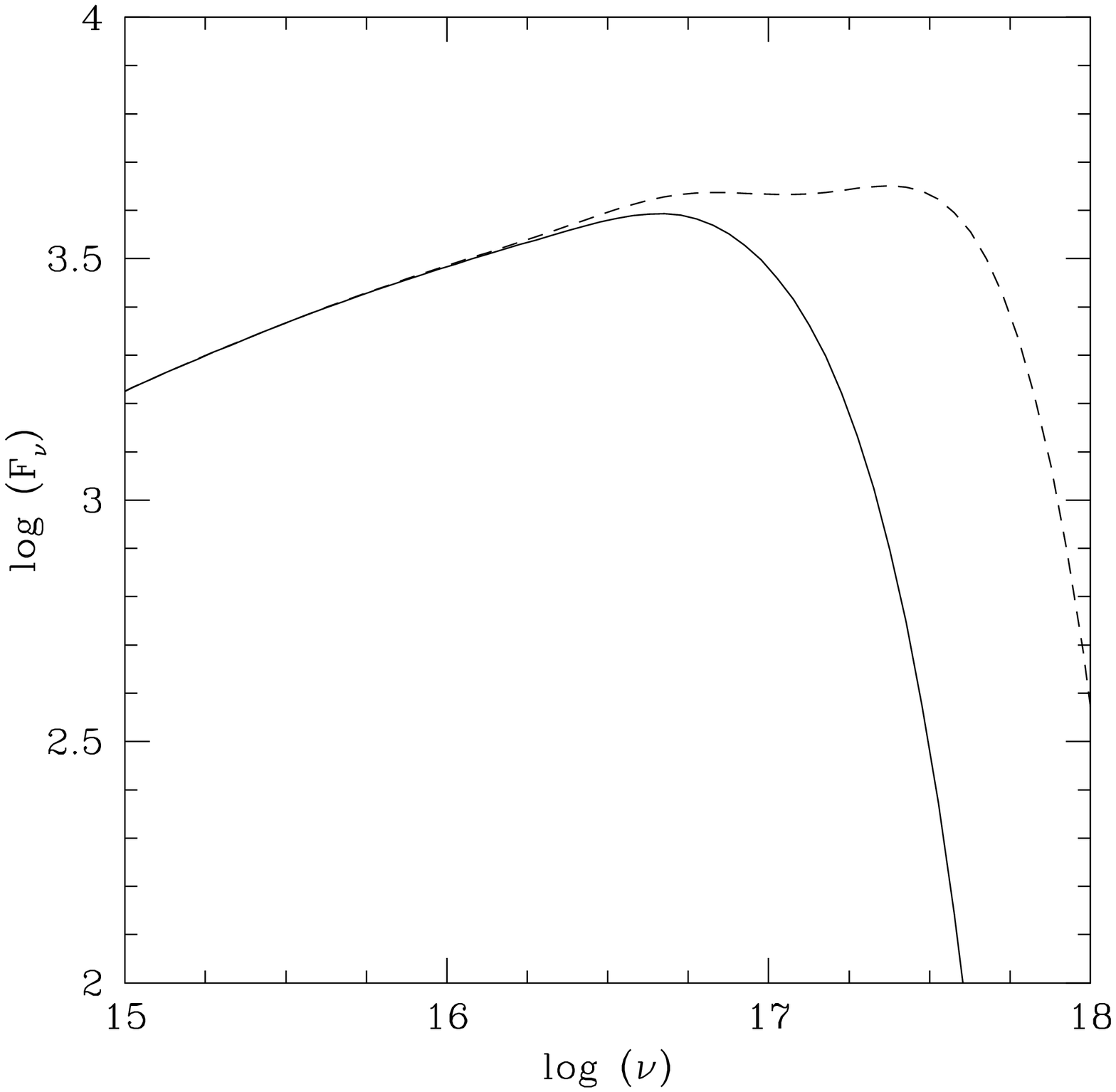}
\caption{The disk blackbody spectrum in the minimum 
(solid line) and maximum (dashed line) luminosity
states.
Model parameters:  $M=10 M_{\odot}$, $\dot m_{\rm ext} = 0.7$,
$\alpha=0.01$. The corona share is $f_{\rm cor}=0.5$,
 but the power law tail is not plotted here. The outflow is neglected. 
\label{fig:spec}}
\end{figure}

In Figure \ref{fig:spec} we show the radiation spectra emitted by
the accretion disk in the maximum (outburst) and minimum luminosity phases. 
We have made a
 simplifying assumption, that
the disk radiates locally as a black body of temperature $T_{\rm eff}(r)$
and the resulting spectrum is integrated over the entire disk. We adopt
the temperature distribution in the unstable region as that computed
from our time dependent model, and we assume that the outside disk is
stable and has a standard temperature distribution 
$T_{\rm eff}\sim r^{-3/4}$.
The spectrum between the outbursts is a standard disk blackbody spectrum.
However, when the disk becomes unstable, the radial temperature
distribution is changed, which affects the spectral slope. Also, the
maximum temperature is much higher, which causes the shift of the
spectrum towards higher frequencies (see Zampieri et al. 2001).
The presence of the corona and/or the outflow should add a power law 
component to the total spectrum, with an extension and slope appropriate 
for the hot plasma properties.
This spectral component is not plotted in the figure.

\section{Discussion}

\subsection{General properties of the model}
\label{sec:general}

The radiation pressure instability operating in the innermost parts of
the accretion disk may be responsible for the periodical 
luminosity changes of a
number of sources. The limit cycle behaviour produces outbursts with
timescales and amplitudes which could reproduce the observed
variability of Galactic black hols (on timescales of
minutes) and short term variability of AGNs (on timescales of years),
provided that certain assumptions about the model are satisfied.

In our time dependent calculations we use vertically averaged
equations of the disk structure. Appropriate numerical coefficients
need to be used in the vertical averaging, because they influence the
critical accretion rate for which the limit cycle starts working. The
description of the disk vertical structure also influences the
computed outburst amplitudes, particularly in the case of the 'disk +
corona' model.

We calculated the time-dependent disk model and associated lightcurves
and spectra assuming that
a moderately strong corona extends above an accretion disk and
contributes to the total accretion energy. Our assumption is
consistent with the observed hard X-ray tail in the spectra and soft
to hard X-ray luminosity ratios (e.g. \.Zycki, Done \& Smith 2001 for
soft X-ray transients, Janiuk, Czerny \& Madejski 2001 for the quasar
PG1211+143).  We also take into account the presence of the vertical outflow.
  These
two assumptions let us obtain good agreement with the observed
GBH lightcurves. For a bare accretion disk without a corona or an outflow,
the outbursts are much too strong in respect to observations and the
maximum luminosity exceeds the Eddington limit.

The outflow efficiency suppresses the maximum luminosity 
during the outburst.  The luminosity reaches even 0.504 $L_{\rm Edd}$ in
the case of no outflow (model with $f_{\rm cor}=0.5$), while for the outflow
strength $A=0.01$ the luminosity drops to $0.2 L_{\rm Edd}$ 
(for the same corona
share).  The total outflow power
could in principle be as high as that emitted in the disk (Falcke \&
Biermann, 1999). Fender (2001) shows that the outflow is likely to be
comparable in power to the integrated X-ray luminosity.
In our calculations the outflow power is variable and depends 
on $L/L_{\rm Edd}$ ratio,
 achieving 30 - 70\% of the observed disk luminosity during the outburst, 
for various sets of parameters.

Not only the outflow but also the corona reduces the outburst amplitude.
 If the fraction
of energy dissipated in corona is small, a strong outflow is needed to
suppress outbursts: in case of $f_{\rm cor}=0.1$ the outflow efficiency
parameter has to be $A > 0.25$.  
For $f_{\rm cor}=0.5$ only a small amplitude flickering 
($\delta log L < 0.05$) occurs when $A > 0.05$. Here we adopt a simplified
description of a corona, assuming constant coronal dissipation
fraction $f_{\rm cor}$, but in general this parameter may depend on the
radius (Janiuk \& Czerny 2000). Recently, Merloni \& Fabian (2002) 
proposed that the coronal fraction may drop for high accretion rates,
being the most efficient in the low state of the source.

\begin{figure}
\plotone{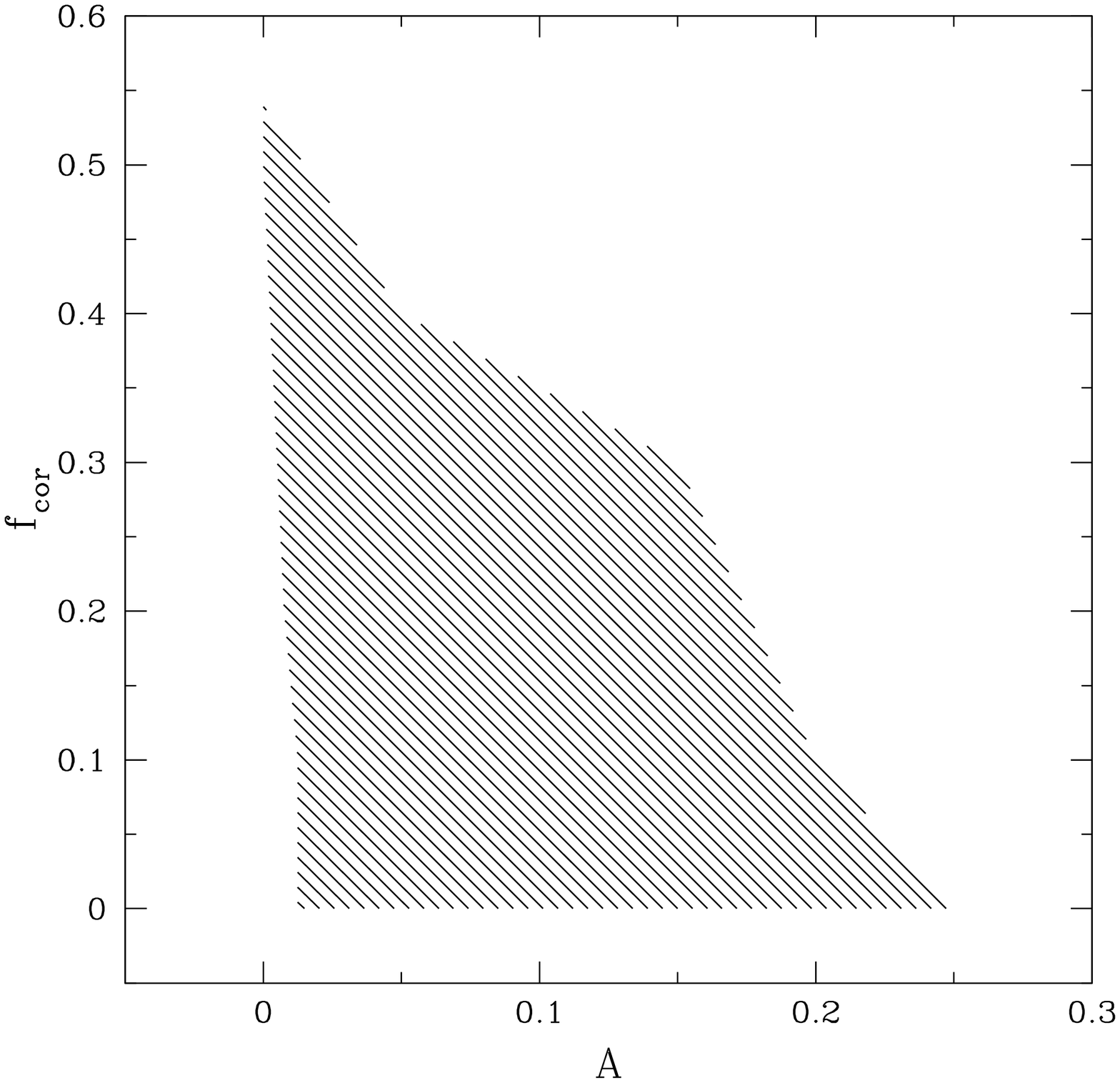}
\caption{Parameter range, for which the outbursts are expected.
Model
parameters: $M=10 M_{\odot}$, $\dot m_{\rm ext} = 0.7$,
$\alpha=0.01$. The shaded region marks the corona and outflow
parameters, for which the luminosity variations are greater than 5\% 
($rms > 0.03$).
\label{fig:topo}}
\end{figure}

When either an outflow or a corona are strong enough ($f_{\rm cor}>0.55$
and/or $A>0.25$), the outbursts do not occur and the disk produces
only flickering. Therefore these two
mechanisms support each other and the stronger the corona the weaker outflow
is required to suppress disk's outbursts.  We illustrate this
correlation in Figure \ref{fig:topo}, where we show on the $A -
f_{\rm cor}$ plane the parameter range for which we detect significant
outbursts.  Here the criterion for outburst detection was chosen that
$rms > 0.03$, that is the mean luminosity variations are above $\sim
5\%$ in logarithm.

The results strongly depend on the adopted external accretion rate. For small
$\dot m_{\rm ext}$ the whole disk remains stable and the outbursts do not
occur. However above a certain critical value the instability develops in the
innermost disk part. 
The critical accretion rate above
which outbursts are expected is $\dot m_{\rm ext}= 0.28$ for a disk
without a corona, but it increases to $\dot m_{\rm ext}= 0.68$ when 
the corona shares 50\% of the energy. 
This critical accretion rate does not depend on whether an outflow is
present or not (see Figure
\ref{fig:sdzet}).
The amplitude of the outburst only weakly depends on the external
accretion rate. 
Instead, the duration of the bursts increases strongly with 
$\dot m_{\rm ext}$, and the 
separation between the outbursts increases as well. For instance, for 
$\dot m_{\rm ext} = 0.56$ the burst duration is $\delta t \sim 250$ seconds, 
while for
 $\dot m_{\rm ext} = 0.39$ it is only $\delta t \sim 150$ seconds 
(model with no corona). 
It is the result of the enlarged instability region 
when the accretion rate is high.  

The outburst amplitude is determined by the separation between $\dot m_{1}$
and $\dot m_{2}$, e.g. the lower and the upper turning points on the {\bf S}
curve (see Section \ref{sec:local}). The point $\dot m_{1}$ is controlled
mainly by whether the corona is present or not and how large is its
contribution to the total accretion energy. The point $\dot m_{2}$ is 
influenced mainly by the strength of the outflow. In order to keep the 
outburst amplitudes no too high and not exceed the Eddington limit in the 
maximum luminosity phase, the point $\dot m_{2}$ must never be too high.
In principle, 
the small value of $\dot m_{2}$  could  alternatively be achieved by 
changing the viscosity
parameter $\alpha$, and would not require any contribution from the hot wind.
 The variable viscosity may be parameterized e.g. by the disk thickness:
$\alpha\sim(h/r)^{-n}$. However, we found that for low viscosities
($\alpha < 10^{-3}$) the unstable radiation pressure dominated branch
becomes too small for the instability to develop, and finally the disk
achieves quasi-stable state in the intermediate range of temperatures
(see  Wallinder 1991).

In our model we assume the $\alpha$ viscosity prescription of 
Shakura \& Sunyaev (1973). If the magneto-rotational instability was taken 
into account, the $\alpha$ parameter would also contain a magnetic 
contribution, which could slightly increase its effective value (Armitage et 
al. 2001). The qualitative results however should not be changed in this case.
On the other hand, we assume a stationary material supply to the radiation 
pressure instability zone from the accretion disk regions outside 
$\sim 200 R_{\rm Schw}$, while in principle additional instability mechanisms, 
such an ionization or gravitational instability, may play crucial role 
further out in the disk (Menou \& Quataert 2001 and references therein).

In all computed cases the duration of the outburst was significantly 
shorter (by a factor of 3 or more) than expected on the basis of the 
viscous timescale computed at the outer edge of the instability strip from
the
analytical formulae provided by Shakura \& Sunyaev (1973). Therefore, 
numerical computations are needed to give appropriate viscous time scale
estimates.  

\subsection{Application to GRS 1915+105}

Observations of the microquasar GRS 1915+105 outbursts imply rather
short time scales and low amplitudes of an outburst.  Feroci et
al. (2001) show lightcurves of the microquasar obtained by Beppo
SAX. Typical soft X-ray band outbursts with $\Delta \log L \sim 0.6$
amplitude last for $\Delta t \sim 100$ sec.  The disk luminosity in
the outburst never exceeds Eddington luminosity. Therefore, based on
our model, the whole limit cycle should run between $\dot m_{1} \sim
0.05$ and $\dot m_{2} \sim 1.0$.

The values of the transition accretion rates $\dot m_1$ and $\dot m_2$
(see Sections \ref{sec:local} and \ref{sec:general}) depend significantly 
on the adopted description of the disk
structure. In particular, the value $\dot m_1$ resulting from the
model of Nayakshin et al. (2000) will be higher by a factor of 5 when the
correct factors computed from the vertical structure model are included. 
It weakens significantly the motivation to postulate a strong coronal
dissipation (up to 90\% in Nayakshin et al. 2000) in this source, 
which is not observed but
which served to increase $\dot m_1$ by a factor of 10. Therefore in
our model we propose a moderately strong corona, dissipating 30-50\%
of the total energy.  If the outflow takes away required amount of the total
emitted power depending on the $L/L_{\rm Edd}$ ratio, the maximum observed
disk luminosity should be lowered.

In our calculations we adopt the black hole mass equal to $10
M_{\odot}$ (recent near infrared spectroscopy results indicate the
mass of $14\pm 4 M_{\odot}$, Greiner 2001) and constant viscosity
parameter $\alpha=0.01$. With a corona dissipating $50\%$ of the total
energy, no outflow, and external accretion rate equal to $70\%$ of
Eddington rate the model results in strong and regular outbursts of
$\Delta log L = 0.7$ amplitude and timescales of $\Delta t \sim 140$
sec.  If the outflow is present than the required by the observations
corona is weaker and dissipates only $30\%$ of energy. In this case
the theoretical lightcurve is quite irregular, with shorter and longer
outbursts lasting for $\Delta t \sim 50 - 100$ sec and variable
amplitudes of $\Delta log L \sim 0.3 - 0.7$.

The scenario of radiation pressure instability has one advantage over
the disk evacuation scenario (Belloni et al. 1997b).  As stressed by
Rao, Yadav \& Paul (2000), the rise/fall timescale during outbursts,
$t \sim 10 s$, is too short for a cold disk to be removed or
reinstalled, as this should happen on a viscous timescale.  In our
model the whole accretion disk exists all the time, just its 
thermal state changes
during outburst, with thermal timescale being short even for
relatively high accretion rates. We see that clearly from Figure 2 of
Janiuk, Czerny \& Siemiginowska (2000) - rise and fall are much 
shorter than the duration of long outbursts typical for high accretion rates.

\begin{figure}
\plotone{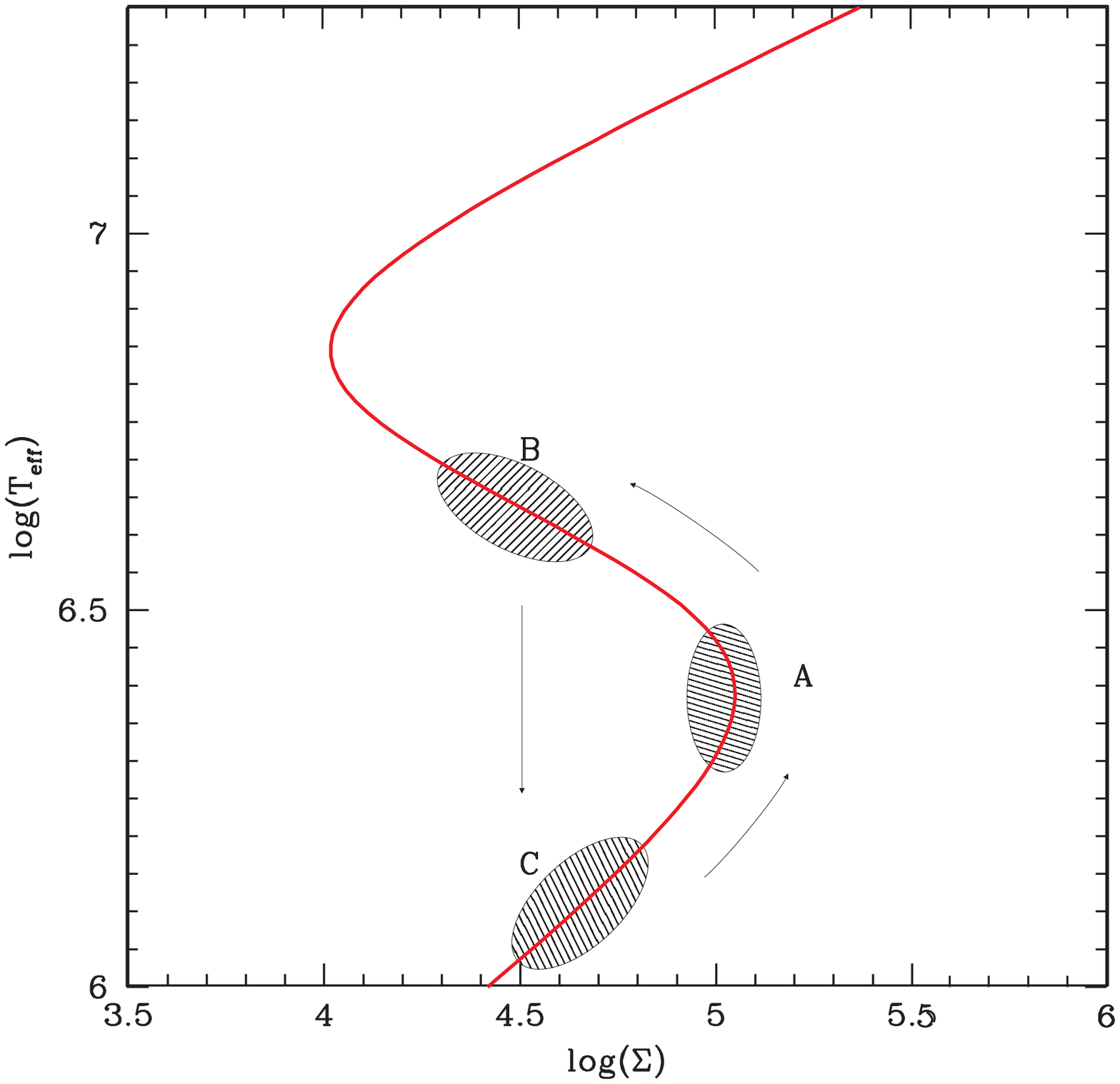}
\caption{ The hardness ratio states A, B, and C, defined as in Belloni et al.
 (2000), identified with the different positions of the disk on the
 stability curve.  The curve is plotted for 10 $R_{\rm Schw}$. The cycle
 performed during the time evolution is marked by arrows.
\label{fig:diagram}}
\end{figure}

Belloni et al. (2000) pointed out, that the source may occupy three
distinct regions on the color - color diagram in X-rays. The colors
are defined as $HR_{1}= F(5-13 keV)/F(2-5 keV)$ and $HR_{2}= F(13-60
keV)/F(2-5 keV)$. The state A corresponds to small values of both
$HR_{1}$ and $HR_{2}$, state B is for $HR_{1} > 1.1$ and small
$HR_{2}$, while state C is for small $HR_{1}$ and large $HR_{2}$.  The
source alterations between states A, B and C are observed
periodically, however the only one way transition between states B and
C has been observed,
namely from B to C, while the reverse transition has never happened.
In our model we could identify the state C with the disk occupying the
low stable branch of the {\bf S} curve (see Figure
\ref{fig:diagram}), with a very soft 
disk blackbody emission and dominating power law component from the
hot corona.  The state A corresponds to the turning point on the
stability curve, when the contribution of the disk blackbody emission
is substantial, and state B refers to the unstable branch on the 
{\bf S} curve. 
In the last case the emission from the hot disk is more pronounced in the 
X-ray spectrum.
The disk normally evolves along the stability curve, going
from C to A and further to B, or back to C state, depending on the
accretion rate. However, when the disk reaches the unstable region B,
the only possibility is to drop down to the state C, when the system
finds the shortest possible way to come back to the stable solution.

\subsection{Luminosity states in other sources and their relation to 
radiation pressure instability}

Galactic black holes (GBH) exhibit a number of distinct luminosity
states (van der Klis 1994, Poutanen, Krolik \& Ryde 1997).  Low (hard)
state is characterized by a power law spectrum, with the photon index
$\Gamma \sim 2$, and a complex variability. 
High (soft) state spectra are dominated by the disk emission, and the
power spectrum is a simple power law with the index $\sim 1$ and a high
frequency cut-off above 100 Hz. When the source gets brighter, in the
Very High State, the spectrum is complex, with considerable and
frequently variable contribution of the power law and disk emission,
and the power spectrum shape is closer to the one in the hard state.
Finally, the faintest sources represent a quiescent state.

Our model based on the radiation pressure instability predicts  
that the luminosity threshold exists and the disk becomes unstable
when the luminosity exceeds this threshold.  There is an important
question to be put forward: are the Low, High and Very High states
related in any way to the radiation pressure instability?

The theoretical luminosity threshold depends on the corona efficiency
and varies from $0.28 L_{\rm Edd}$ for no corona ($f_{\rm cor}=0$) to $0.68
L_{\rm Edd}$ for $ f_{\rm cor}=0.5$.  The range of observed accretion rate
corresponding to various luminosity states is rather difficult to
assess. Firstly, only the luminosity in a restricted energy band is
usually available, while estimates of the bolometric luminosity are
reliable only if the data truly represent the broad spectrum.
Secondly, there may sometimes be a confusion in the state definitions.
The Very High State looks similar to both the hard state and the soft
state, and it may be classified as one of those two, unless a careful
analysis of both the spectral and temporal behaviour is
performed. Keeping this in mind, we have collected a number of
accretion rate estimates from the literature for the purpose of
qualitative discussion.  In Table~\ref{tab1} we show the estimates of
the accretion rate (or luminosity to the Eddington luminosity
ratio). Information is not complete since many states are observed
mostly in transient sources and then the evolution may be too fast to
collect many data (or sometimes even a single data set) for a given
luminosity state.

There seem to be a large difference between the range of accretion
rates in different states among various objects. Systematic errors and
dispersion in inclination angles may partially be responsible for
those observed differences, but there may also be some additional
parameters hidden in the description of formation of the hot
(coronal?) plasma. Still, sources are in their Low States when $\dot m
\sim 0.05$, in the High States when $\dot m \sim 0.1$, and in the Very
High States when $\dot m \sim 0.5$.  The luminosity range
corresponding to the High state is quite narrow, factor of 2 -- 3 in
accretion rate between the minimum and maximum value.

The Low and the High states are basically expected to be
stable. The difference between these states does not seem to be
related to radiation pressure instability but there must be a
mechanism of hot plasma generation which works less efficiently when
the total flux is larger. There are several phenomenological models
proposed but the physics of formation and geometry of hot plasma is
under discussion (e.g. Poutanen 2001, Di Salvo et al. 2001).  Some
objects in the High State are possibly slightly above the threshold
predicted by our model although their coronal emission is not stronger
than $f_{\rm cor}=0.5$ (LMC X-3 and J1748-288). It should be considered
whether these sources are indeed in the High State, or rather in a
disk dominated Very High State.

Sources in their Very High State are definitively in the range
corresponding to the radiation pressure instability. Outbursts lasting
a hundred to a thousand of seconds are expected. Such outbursts are
however seen only in microquasars - GRS 1915+105 (see Sect. 4.2) and
GRO J1655-40 (Remillard et al. 1999). The lack of outbursts in other
sources is consistent with our model only if the corona and/or outflow
are strong enough to suppress the instability completely. Since the
'corona' (observationally - a power law component) in the Very High State
contributes typically to about 30 \% of energy, considerable outflow is
also needed (parameter $A$ in our model of order of 0.2).  In
sources, which do not exhibit jets, the outflow must be
uncollimated. The model does not include any description of the jet
formation process, therefore it basically applies to such uncollimated
outflow.

It remains to be checked whether this outflow is consistent with
observations.  Energetically, the request for such an outflow is plausible 
since the observed collimated jet may carry a considerable fraction 
of a total energy (Fender et al. 1999, Reeves et al. 2001). However, 
this material, if uncollimated, is in our line of sight
and covers the disk, so part of the 'corona' effect in the Very High
State may be due to the Comptonization in this outflowing
material. Such winds were also discussed in several contexts
(e.g. wind driven instability of Shields et al. 1986, Beloborodov
1999, ADIOS models of Blandford and Begelman 1999). In the present
model the wind should develop as {\it a result} of the disk
instability, thus providing a stabilizing mechanism, and therefore it
should cover only the radiatively unstable part of the disk. This hot
wind plasma should have different properties than the hot coronal
plasma formed in the Low state, which is probably produced by
another mechanism.

\begin{deluxetable}{lcccr}
\tablenum{1}
\tablewidth{160 mm}
\tablecaption{Luminosity ranges in Eddington units
\label{tab1}}
\tablehead{
\colhead{Object} &
\colhead{Very High State} &
\colhead{High State} &
\colhead{Low State} &
\colhead{References} 
}
\startdata
 J1748-288   & 0.30 - 0.50  & 0.05 - 0.25  & 0.01 - 0.04 & Revnivtsev et al. 2000\\
 Nova~Muscae & 0.09 - 0.50  & 0.03 - 0.09  & 0.006       & \.Zycki et al. 1998\\
 LMC X-3     & $-^{b}$      & 0.07 - 0.3   & $< 0.07$    & Wilms et al. 2001\\
 Cyg X-1     & $-^{b}$      & $> 0.03$     & $< 0.03$    & Gierli\' nski et al. 1997, \\
             &              &              &             &Gierli\' nski et al. 1999\\      
 GS 2023     & $< 1.6$      & 0.07         & 0.04 - 0.05 & \.Zycki et al. 1999 a,b\\      
 GRO J1615   & 0.2 - 0.25   & 0.1          & $-^{a}$     & Remillard et al. 1999,\\
             &		    &		   &		 & Gierli\' nski et al. 2001 \\      
 LMC X-1     & $-^{b}$      & 0.15         & $-^{b}$     & Gierli\' nski et al. 2001\\      
 GX 339-4    & 0.5          &  $-^{a}$     & 0.05 - 0.15 & Miyamoto et al. 1995,\\
	     &		    &		   &		 & Wilms et al. 1999\\      
 GRS 1915+105& $>0.15$      &  $-^{b}$     & $-^{b}$     & Zdziarski et al. 2001,\\
	     &		    &		   &		 &Belloni 1997b\\ 
 	     &		    &		   &		 &Trudolyubov et al. 1999 \\
    
\enddata

\tablenotetext{a}{
No observations in this state}
\tablenotetext{b}{
No transitions to this state}
 
\end{deluxetable}

\subsection{The vertical outflow}

In this work we postulate that a certain amount of the accretion energy is 
taken away by the vertical outflow. Since we do not include any collimation 
mechanism, this outflow has a form of a hot wind rather than a collimated jet,
however its origin are the innermost regions of the accretion flow.
We did not study this outflow in detail, since we took into account only the 
energy loss due to the wind. We neglected the mass loss and the effect 
of the hot flow onto the disk spectrum. However, we can estimate the properties of the outflowing wind a posteriori.
The energy flux of the outflowing material is given by:

\begin{equation}
F_{\rm out}={1 \over 2} \dot m_{\rm z} v_{\rm z}^{2},
\end{equation}  
where $v_{\rm z}$ is the outflow velocity in the vertical direction and 
$\dot m_{\rm z}$ is the mass loss rate. Therefore the outflow efficiency, 
defined by the Equation (\ref{eq:fjet}), is equal to:

\begin{equation}
f_{\rm out}={F_{\rm out} \over F_{\rm tot}}={4 \pi r^{2} \dot m_{\rm z} \over 3 \dot M f(r)}
\times \big({v_{\rm z}\over v_{\rm \varphi}}\big)^{2},
\end{equation}
where local flux $F_{\rm tot}$ is given by Equation (\ref{eq:ftot})
 and $v_{\rm \varphi}$
is the Keplerian velocity on the orbit $r$.
Therefore the mass loss rate will be given by:
\begin{equation}
\dot m_{\rm z} = {\dot M_{\rm Edd} A \dot m^{3} \over 4 \pi r^{2}(1+A \dot m^{2})} 
\times \big({v_{\rm \varphi}\over v_{\rm z}}\big)^{2} 3 f(r)
\end{equation}
and may be negligible if e.g. $v_{\rm z}>3 v_{\rm \varphi}$.

For the exemplary model with $A=0.1$, we set $\dot m \sim 1$ and 
$(v_{\rm \varphi}/v_{\rm z})^{2}\sim 0.1$ at $r=10 R_{\rm Schw}$. The estimated density 
of the wind is then equal to $\rho_{\rm wind}=\dot m_{\rm z}/v_{\rm z}=2.7\times 10^{-9}$ 
g cm$^{-3}$. This means that the optical depth is of the order of 
$\tau_{\rm es} \sim 0.03$ and the outflowing medium should not have any 
substantial impact on the observed spectral properties of the X-ray source. 
The Comptonization in the wind may lead at most to the 
formation of a weak, steep 
power-law tail, additional to the disk black body emission. 
The X-ray reflection amplitude is also reduced by the bulk motion in the range
 of $\beta_{\rm bulk} \sim 0.6 - 0.7$, so this effect is practically 
negligible (Janiuk, Czerny \& \. Zycki 2000).

We stress that the uncollimated outflow which partially stabilizes the 
radiation pressure dominated disk is launched by a different mechanism than 
the jet seen in GRS 1915+105 during the prolonged C states. The latter is 
observed in the low luminosity intervals and is probably connected with the
 radio emission  (Klein-Wolt et al. 2001).

\section{Conclusions}

Time dependent evolution of the inner accretion disk is 
based on the radiation pressure instability, playing major role in so called 
'$\alpha$ disks'. This process accounts for a number of 
properties of Galactic black hole binary systems.

\begin{itemize}
\item The radiation pressure dominated innermost regions of the 
accretion disk are unstable and may oscillate between cold and hot phase. 
When the hot front goes through the inner disk, its surface density drops 
down rapidly by a factor of 3. 
As the disk cools down, it switches to the state of a relative faintness
and the X-ray spectrum may be dominated by a power law emission from 
a hot corona.
However, the disk is never evacuated 
completely and extends always down to the marginally stable orbit 
around a black hole.   
\item For high enough accretion rates ($\dot m > 0.28$)
and moderate strengths of the outflow and the corona, the instability 
results in a characteristic variability pattern, with
100 - 200 sec. outbursts of an amplitude reaching 
$\Delta log L \sim 0.3 - 0.7$. The irregular outbursts' shape, observed 
in the microquasar GRS 1915+105, may result from the presence of the
outflow, which takes away a high fraction of the energy when the disk 
luminosity is close to the Eddington limit.  
\item The presence of luminous outbursts and their amplitude depends on 
the assumed strengths of the disk corona and the outflow.  These two
elements may weaken the outbursts or even suppress them completely,
therefore in most systems accreting at high $\dot m$ (observed in the
Very High State) a rapid, large amplitude variability is not detected.
\item The distinct regions occupied periodically by the microquasar 
on the X-ray color-color diagram correspond in our model 
to the characteristic phases during the time evolution.
The theoretically predicted evolutionary track through these states: $C \rightarrow A \rightarrow B \rightarrow C$ explains the lack of  $C \rightarrow B$ 
transitions. Observed transitions $B \rightarrow A$ and $A \rightarrow C$
and the fast X-ray flickering are not accommodated yet in the model, since 
they are both probably caused by hot plasma instabilities.

\end{itemize}

{\sl Acknowledgments.}  We thank Piotr \. Zycki for helpful
discussions. This work was supported in part by grant 2P03D00322 of
the Polish State Committee for Scientific Research. AJ acknowledges
support of the Harvard - Smithsonian Center for Astrophysics. 
This research was funded in part by National Aeronautics and 
Space Administration 
through Chandra Award Number GO1-2117B issued by the Chandra
X-ray Observatory Center, which is operated by Smithsonian Astrophysical
Observatory for and on behalf of the NASA under contract NAS8-39073.


\end{document}